\documentclass[preprint]{aastex}
\usepackage{apjfonts}
\usepackage{amsmath}
\usepackage{graphicx}
\usepackage{natbib}
\usepackage{longtable}
\usepackage{booktabs}
\shorttitle{Failed Kepler multis}
\shortauthors{Hwang et. al.}

\def\pasa{\rm{PASA}}

\begin{document}

\title{Dynamics and Collisional Evolution of Closely Packed Planetary Systems}

\author{Jason A. Hwang$^{1,2}$, Jason H. Steffen,$^{1,2,3}$ J. C. Lombardi Jr.,$^4$ \& Frederic A. Rasio$^{1,2}$}
\affil{$^1$Northwestern University, Department of Physics and Astronomy, Northwestern University, 2145
  Sheridan Road, Evanston, IL 60208, USA}
\affil{$^2$Center for Interdisciplinary Exploration and Research in Astrophysics (CIERA), Northwestern University, 2145
  Sheridan Road, Evanston, IL 60208, USA}
\affil{$^3$Department of Physics and Astronomy, University of Nevada, Las Vegas, 4505 S. Maryland Pkwy., Las Vegas, NV 89154-4002, USA}
\affil{$^4$Department of Physics, Allegheny College, Meadville, PA 16335, USA}

\begin{abstract}
High-multiplicity Kepler systems (referred to as Kepler multis) are often tightly packed and may be on the verge of instability.
Many systems of this type could have experienced past instabilities, where the compact orbits and often low densities make physical collisions likely outcomes.
We use numerical simulations to study the dynamical instabilities and planet-planet interactions in a synthetically generated sample of closely-packed, high-multiplicity systems.
We focus specifically on systems resembling Kepler-11, a Kepler multi with six planets, and run a suite of dynamical integrations, sampling the initial orbital parameters around the nominal values reported in \citet{2011Natur.470...53L}, finding that most of the realizations are unstable, resulting in orbit crossings and, eventually, collisions and mergers.
We study in detail the dependence of stability on the orbital parameters of the planets and planet-pair characteristics to identify possible precursors to instability, compare the systems that emerge from dynamical instabilities to the observed Kepler sample (after applying observational corrections), and propose possible observable signatures of these instabilities.
We examine the characteristics of each planet-planet collision, categorizing collisions by the degree of contact and collision energy, and find that grazing collisions are more common than direct impacts.
Since the structure of many planets found in Kepler multis is such that the mass is dominated by a rocky core, but the volume is dominated by a low-density gaseous envelope, the sticky-sphere approximation may not be valid, and we present hydrodynamic calculations of planet-planet collisions clearly deviating from this approximation.
Finally, we rerun a subset of our dynamical calculations using instead a modified prescription to handle collisions, finding, in general, higher multiplicity remnant systems.
\end{abstract}


\section{Introduction}
\label{Sec:Intro}
Of the thousands of planetary candidates discovered by the {\it Kepler} mission (\citealt{2015ApJS..217...31M}; \citealt{2014ApJ...784...45R}; \citealt{2013ApJS..204...24B}; \citealt{2010Sci...327..977B}), roughly a third are in known multiple transiting-planet systems.
While a gas or planetesimal disk tends to damp the eccentricities and inclinations of the planets in a system---generally driving the system into more stable orbits, there is no known guarantee that a system remains stable after the disk dissipates.
\citet{2013ApJ...767..115F} show that many multiple-planet systems are likely to be 'dynamically packed', meaning that there are few stable orbits for additional planets \citep{2004ApJ...611..494B}, and fit the the underlying dynamical-separation distribution as a shifted Rayleigh distribution with a mean dynamical separation of $\Delta=21.7$ and a Rayleigh parameter $\sigma=9.5$, where the dynamical separation is defined as \begin{equation}\label{EQ:DynamicalSeparation}\Delta\equiv\frac{a_2-a_1}{R_H},\end{equation} where $R_H$ is the planets' mutual Hill radius, defined as \begin{equation}R_\mathrm{H}=\left(\frac{m_1+m_2}{3M_*}\right)^{1/3}\left(\frac{a_1+a_2}{2}\right),\end{equation} where $a_1$ and $a_2$ are the semi-major axes of the inner and outer planets, $m_1$ and $m_2$ are masses of the inner and outer planets, and $M_*$ is the mass of the star.
\citet{2011ApJS..197....8L} studies the characteristics of multiple-planet systems in detail, noting that there are many short-period, nearly coplanar, multiple-planet systems and that most of the planets in these systems have radii between $1.5-6.0\ R_\oplus$.
Given the number of dynamically-packed systems, many of these could have undergone dynamical instability, resulting in planet-planet scattering and physical collisions.
\citet{2015ApJ...806L..26V} conduct dynamical studies on the stability of these tightly-packed systems and suggest that nearly all stars began with higher-multiplicity planetary systems, and that the currently observed population is a remnant of dynamical instabilities.
\citet{2012ApJ...755L..21D} demonstrates that the Kepler-36 system almost certainly undergoes short-timescale dynamical chaos, and while this chaos does not necessarily result in orbit crossing, Kepler-36 is a good example of a system where small changes to the orbital elements produce a material change in the dynamical evolution.
Similarly, there are other exoplanet systems where small deviations from the observed orbital elements would lead to instability (e.g. GJ-876 and HD 82943, \citealt{2004ApJ...617..569B}).
Understanding the architecture of the remnants after such planet-planet interactions may provide clues in tracing the collisional history of the currently observed Kepler systems.
Studies of the orbital architectures of planetary systems (e.g., \citealt{2015ApJ...807...44P}; \citealt{2015MNRAS.448.1956S}; \citealt{2011ApJS..197....8L}; \citealt{2012ApJ...750..114F}; \citealt{2013MNRAS.433.3246S}) may be used to gain a statistical understanding of the fraction of systems that may have undergone post-disk dynamical instability.
Our goal is to study the outcomes of such instabilities, including detailed 3-D hydrodynamic calculations of collisions.
In this work we focus on systems like Kepler-11, using the nominal masses and orbital parameters reported in \citet{2011Natur.470...53L} to create many realizations of systems on the verge of instability as a first look at the dynamics of planet-planet interactions in high-multiplicity systems.

Many studies have been conducted on the stability of multiple-planet systems, beginning with circular and coplanar 2-planet systems and more recently, high multiplicity, non-circular, and non-coplanar systems.
The stability criterion for a two-planet system with circular and coplanar orbits is \begin{equation}\Delta>2\sqrt{3}\simeq3.46.\end{equation}
This condition ensures that orbits cannot cross and is known as 'Hill Stable' (see \citealt{1993Icar..106..247G} for generalizations).
\citet{1996Icar..119..261C} demonstrates that all systems with three or more planets eventually become unstable with an instability timescale \begin{equation}\label{EQ:t_instability}\log(t_\mathrm{c}/t_\mathrm{0})=b\Delta+c,\end{equation} where $t_\mathrm{c}$ is the time to the first planet-planet interaction, $t_\mathrm{0}$ is a characteristic timescale, and $b$ and $c$ are constants dependent on the multiplicity and planet masses.
\citet{2011ApJS..197....8L} find that the majority of the planet pairs in the Kepler sample fulfill the 'Hill Stable' requirement and are dynamically stable for more than $10^{10}$ orbits of the innermost planet based on numerical integrations.
\citet{2009Icar..201..381S} find that larger planet masses, higher multiplicities, and tighter spacings result in lower instability times for systems with equal-mass and equally-spaced planets.
Specifically, for systems with $\Delta<3.4$ and an innermost planet at $1$ AU, the time to first orbit crossing is roughly constant at about $10$ years, for systems with $3.4\le\Delta\le8.4$, the time to instability follows \eqref{EQ:t_instability}, and for systems with $\Delta>8.4$, the time to instability increases much more sharply with $\Delta$.
Finally, they report that in a system of terrestrial-like planets, the addition of a high mass planet with a semi-major axis much larger than the outermost terrestrial planet has a small, destabilizing effect on the system.
\citet{2015ApJ...807...44P} study the stability of non-circular and non-coplanar seven-planet systems and find that non-circular and non-coplanar systems require a higher dynamical separations to remain stable for a given timescale.
\citet{2015ApJ...806L..26V} integrate several analogs of tightly-packed systems, including Kepler-11, and characterize the instabilities that occur, hypothesizing that nearly all planetary systems began as a tightly-packed, higher-multiplicity system.

We generate a synthetic population of systems from the orbital parameters reported for Kepler-11 in \citet{2011Natur.470...53L}, a tightly-packed system with six planets, sampling the semi-major axes, eccentricities, and masses within the nominal error-bars.
Kepler-11b and Kepler-11c have one of the smallest dynamical separations, $\Delta=4.7$, of all planet pairs and one of the smallest period ratios of non-resonant planet pairs in the Kepler sample \citep{2011Natur.470...53L}.
We study in detail the influence of the chosen initial conditions on the stability of the system, specifically the orbital parameters of the planets and the characteristics of each planet pair.
We observe that $\sim92\%$ of the systems are unstable, exhibiting orbit crossing and planet-planet collisions, and compare the remnant systems to the currently observed Kepler sample, correcting for observational biases.
We also examine the characteristics of the collisions that occur and find a preference for grazing impacts, where only the gas envelopes come into contact.

The outcomes of these planet-planet interactions depend on many factors including the ratio of the escape velocity from the planet's surface to the orbital escape velocity; specifically, instability in systems with giant planets will often lead to planetary ejections (\citealt{1996Sci...274..954R}; \citealt{2008ApJ...686..580C}), while compact systems with smaller planets, such as Kepler-11, are less capable of producing ejections and dynamical instability more frequently leads to planet-planet collisions \citep{2008ApJ...686..621F}.
Many studies have been conducted of potential Earth-Moon forming collisions, specifically between a proto-earth and an impactor, each with various iron-core to silicate-mantle mass ratios (\citealt{1986Icar...66..515B}; \citealt{2000orem.book..133C}; \citealt{2001Natur.412..708C}; \citealt{2004ARA&A..42..441C}; \citealt{2013Icar..222..200C}; \citealt{2014DPS....4650109C}) and using various tabulated equations of state to handle the abrupt changes in density and phase transitions of the impacted rock.
\citet{2015ApJ...812..164L} present calculations of direct collisions between a rocky planet and a gaseous planet, treating the rocky core with a multi-phase equation of state (Tillotson 1962) and the gas as a polytrope with $\gamma=5/3$.

Since the typical planets found in Kepler multis have masses $\approx3.0M_\oplus-9.0M_\oplus$ (\citealt{2015ApJ...806..183W}; \citealt{2014ApJ...787...80H}; \citealt{2014ApJ...783L...6W}; \citealt{2011ApJS..197....8L}) dominated by a rocky core, but volumes dominated by a low-density, gaseous atmosphere, the qualitative outcomes of collisions, specifically if the planets merge and the fraction of the atmosphere retained, depend sensitively on the distance of closest approach and kinetic energy of the collision \citep{2010ApJ...714L..21K}.
The structure of sub-Neptunes combined with the preference for grazing collisions motivates us to study these types of collisions to improve upon the sticky-sphere approximation used in many N-body integrators.
We present two representative hydrodynamic calculations of collisions between Kepler-11d and Kepler-11e from our suite of dynamical integrations, where we use the gas-mass fractions reported in \citet{2013ApJ...770..131L}.
We use detailed models, generating the gas envelope using {\it Modules for Experiments in Stellar Astrophysics} (MESA) (\citealt{2011ApJS..192....3P}; \citealt{2013ApJS..208....4P}; \citealt{2015ApJS..220...15P}) and the core by integrating backwards from the envelope-core interface using the semi-analytic polytropic equations of state from \citet{2007ApJ...669.1279S} (see \S\ref{Sec:GasEnvelope} and \S\ref{Sec:DiffCore} for more details).
The treatment of sub-Neptunes presents unique challenges as the inner-core, silicate mantle, and gas envelope are each sizable and must be modeled in detail, leading to density discontinuities that are difficult to resolve and integrate with Smoothed-Particle Hydrodynamics (SPH).
\S\ref{Sec:EoS} describes our treatment of these discontinuities by introducing mixed-composition particles to better resolve the interfaces between different materials, such as the core-mantle and mantle-gas boundaries.
To study the structure of merger remnants we use a prescription, utilizing models from \citet{2014ApJ...792....1L}, to predict observable properties of mergers, specifically to identify possible observational signatures of prior instabilities.
Finally, we rerun a subset of our dynamical integrations, using a modified prescription to treat collisions, to study the effects of relaxing the sticky-sphere approximation, and find, in general, higher multiplicity remnant systems.

The rest of the paper is organized as follows: In \S\ref{Sec:Methods} we discuss our choice of initial conditions, specifically how we generate our suite of realizations and the numerical methods used in the N-body code.
In \S\ref{Sec:Results} we present the results of the dynamical integrations, discussing stability timescales, the influence of the chosen initial conditions on the stability of the system, and the architecture of the remnant systems compared to the observed Kepler sample.
In \S\ref{Sec:Collisions} we investigate the characteristics and orbital evolution of collisions seen in the realizations, discuss the potential outcomes, and present possible observational signatures of mergers using a prescription from \citet{2014ApJ...792....1L}.
In \S\ref{Sec:Consequences} we present our investigation into the accuracy and implications of using the sticky-sphere approximation, including two SPH calculations of collisions occurring in our Kepler-11 realizations.
Finally, we summarize our conclusions in \S\ref{Sec:Conclusions}.

\section{Discussion of Parameter Space and Methods}
\label{Sec:Methods}
We are primarily interested in the dynamical evolution of high-multiplicity, closely-packed systems that undergo instabilities, specifically the details of the planet-planet interactions and the architecture of remnant systems.
Kepler-11 provides a nominal example of a high-multiplicity, closely-packed system likely on the verge of dynamical instability; the inner five planets have small dynamical separations, with $5<\Delta<15$, compared to most other high-multiplicity systems detected by Kepler.
While more biased than generating a generic set of initial conditions, our choice of sampling from the nominal Kepler-11 system (within the error bars) has many advantages.
Most importantly, we avoid introducing artificial structure in the orbital separations between planets (e.g. compared to a simple distribution of initial periods or dynamical separations in constant ratio).
Additionally, our procedure naturally approximates the actual scalings present in real systems (for example, inner planets have lower masses and gas-mass fractions).
Conveniently, this choice of initial conditions also automatically results in a high fraction of systems that become unstable on an appropriately long timescale ($\sim10^6$ orbits of the innermost planet; see \S\ref{SSec:Stability_Timescales}), allowing us to focus on the outcomes of this subset of systems.

We use a Monte Carlo method to generate initial conditions using the nominal values from the all-eccentric fit reported in supplementary table 2 from \citet{2011Natur.470...53L}.
We sample the masses, semi-major axes, and eccentricities from a normal distribution generated from the reported mean and standard deviation, resampling negative parameters.
We sample the inclination using the best fit model suggested by \citet{2012ApJ...761...92F}, a Rayleigh distribution with a Rayleigh parameter $\sigma=1.0^\circ$, consistent with previous studies of the underlying distribution describing the coplanarity of Kepler multis (\citealt{2011ApJS..197....8L}; \citealt{2012AJ....143...94T}; \citealt{2014ApJ...790..146F}).
We use the reported nominal values for the radius of each planet and scale the density with the assigned mass.
Since the mass and orbital elements, other than semi-major axis, for Kepler-11g are poorly constrained, we estimate $e\cos\omega$ and $e\sin\omega$ simply as an average of the other planets' values, the mass using a linear fit of the radius, interpolating between Kepler-11d and Kepler-11e, and the density as the average density scaled appropriately to the mass and radii of the other planets.
Because the dynamical separation between Kepler-11f and Kepler-11g is much larger than the other planet pairs ($\Delta=25$), the rough estimate of the mass should have minimal effect on the dynamical stability.
Finally, we use the reported host star's mass $M_*=0.95\ M_\odot$ and radius $R_*=1.0\ R_\odot$ \citep{2011ApJS..197....8L}.
While we use a single system to generate our suite of dynamical calculations, we are able to sample thoroughly both orbital elements of individual planets and planet-pair characteristics (see \ref{SSec:Stability_Initial_Conditions} for details).

We exclusively use the Burlish-Stoer integrator within the N-body dynamics package, {\it Mercury} 6.2 \citep{1999MNRAS.304..793C}, for our integrations.
We find significant improvement in energy conservation using the Burlish-Stoer integrator over both the Symplectic and Hybrid integrators, likely due to the number of close encounters in our dynamical integrations.
We integrate each system to a maximum of $100$ Myr, more than $10^{9}$ orbits of the innermost planet, with an accuracy parameter $\epsilon=10^{-12}$.
In the primary suite of integrations we treat physical collisions in the simple sticky-sphere limit and rerun a subset using a modified prescription for collisions to investigate deviations from this approximation (see \S\ref{Sec:NBody2}).

\section{Results from Dynamical Integrations}
\label{Sec:Results}

\subsection{Stability Timescales}
\label{SSec:Stability_Timescales}
Here we present the results of the dynamical integrations, removing from the results any systems with an initial dynamical separation that is not 'Hill Stable' ($\Delta>2\sqrt3$).
From our remaining sample of 694 dynamical integrations, we find that only 54 ($8\%$) remain stable on the order of $10^{9}$ orbital periods of the innermost planet, 130 ($19\%$) result in a 5-planet system, 306 ($44\%$) result in a 4-planet system, 197 ($28\%$) result in a 3-planet system, and 7 ($1\%$) result in a 2-planet system.
In total we observed 1361 planet-planet collisions.

\begin{figure}[htp]
\begin{center}
\begin{tabular}{cc}
\includegraphics[width=10.0cm]{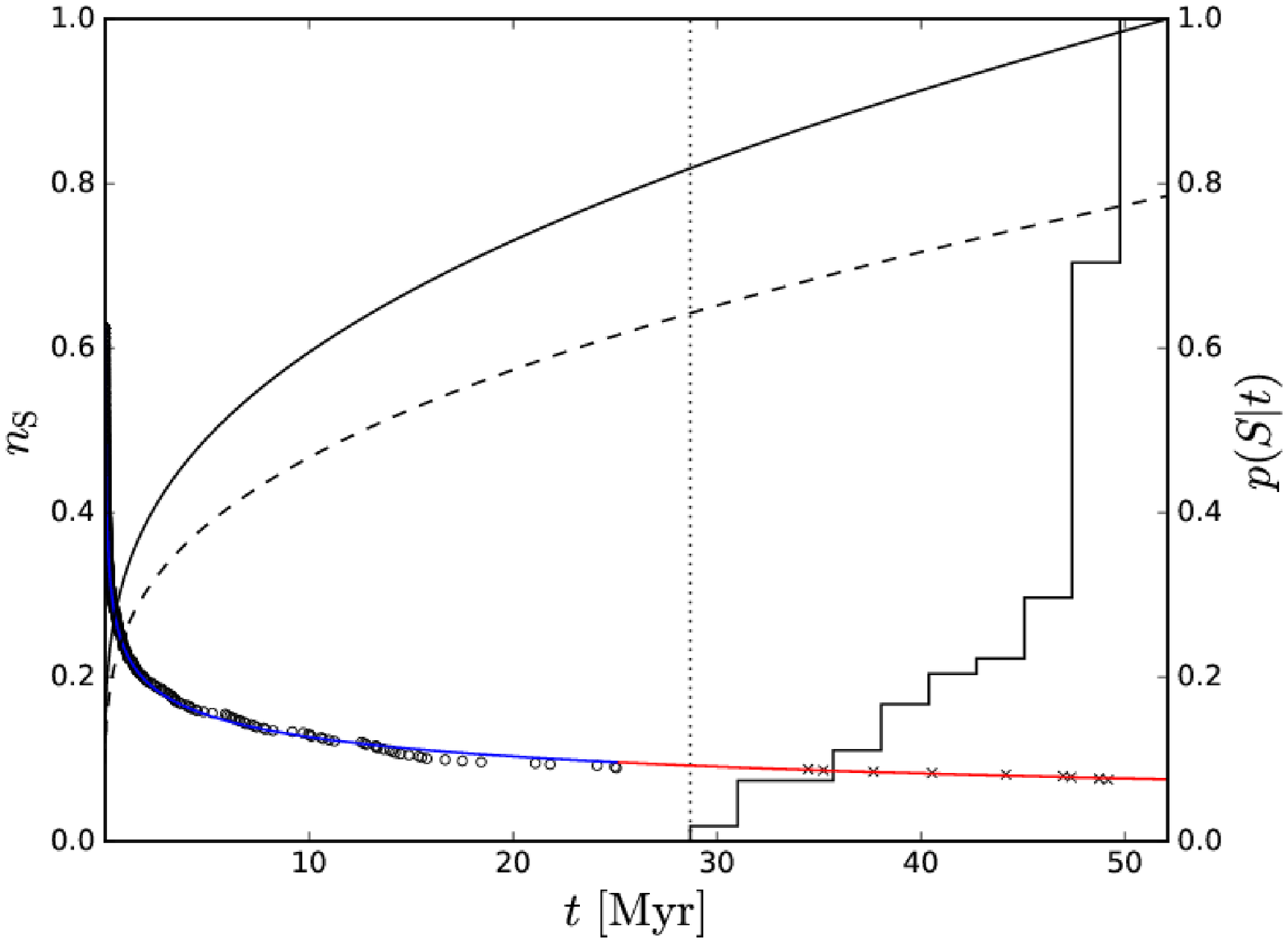}\\
\includegraphics[width=10.0cm]{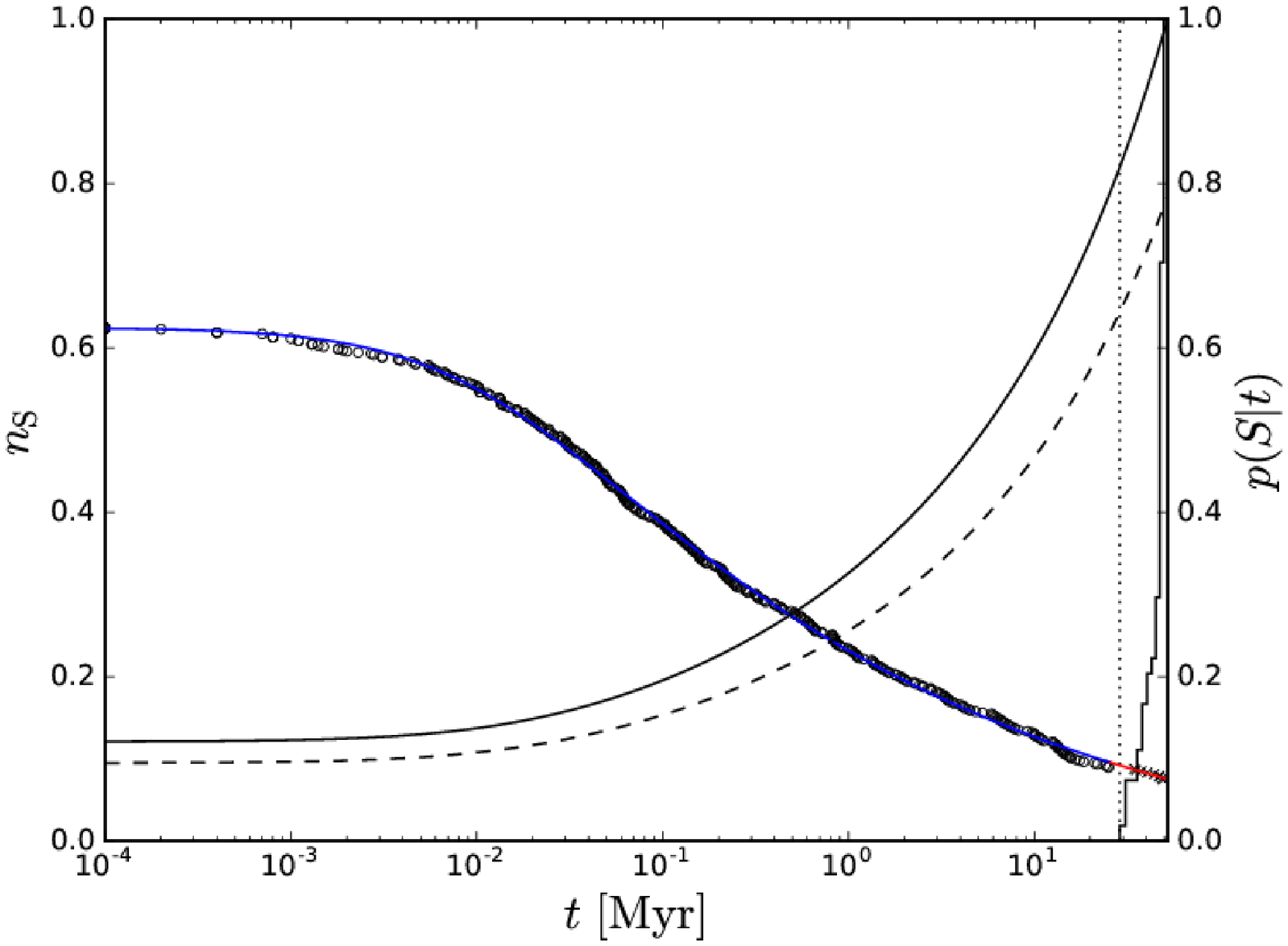}\\
\end{tabular}
\end{center}
\caption{Fraction of systems that are stable as a function of integration time on a linear (top) and logarithmic (bottom) scale.
The histogram shows the distribution of integration times of stable systems; all integrations have reached at least $t_\mathrm{min}=28.7$ Myr (vertical dashed line).
We fit the fraction of systems that are stable with a power law \eqref{EQ:PowerLaw} up to $t_\mathrm{min}$ (blue) and mark the occurrences of first instabilities and the corresponding fraction of stable systems (circles).
The first circle at $t=10^{-4}$ Myr shows the fraction of systems that are stable at the first time step; a significant fraction of systems experience an instability at $t<10^{-4}$ Myr.
We fit the fraction of systems that are stable with an extrapolated power law for $t>t_\mathrm{min}$ (red) and mark the projected fraction of systems that are stable ($\times$, see \S\ref{A:Bayes}).
We show the probabilities that the system is stable until $52\ \mathrm{Myr}$ (solid line) and $100\ \mathrm{Myr}$ (dashed line), given that the system is stable at time $t$,
and find $96\%$ of the stable systems will remain stable until $52\ \mathrm{Myr}$, roughly $10^9$ orbits of the initial innermost planet.}
\label{Fig:t_integrations}
\end{figure}

\begin{figure}[htp]
\begin{center}
\begin{tabular}{cc}
\includegraphics[width=8.0cm]{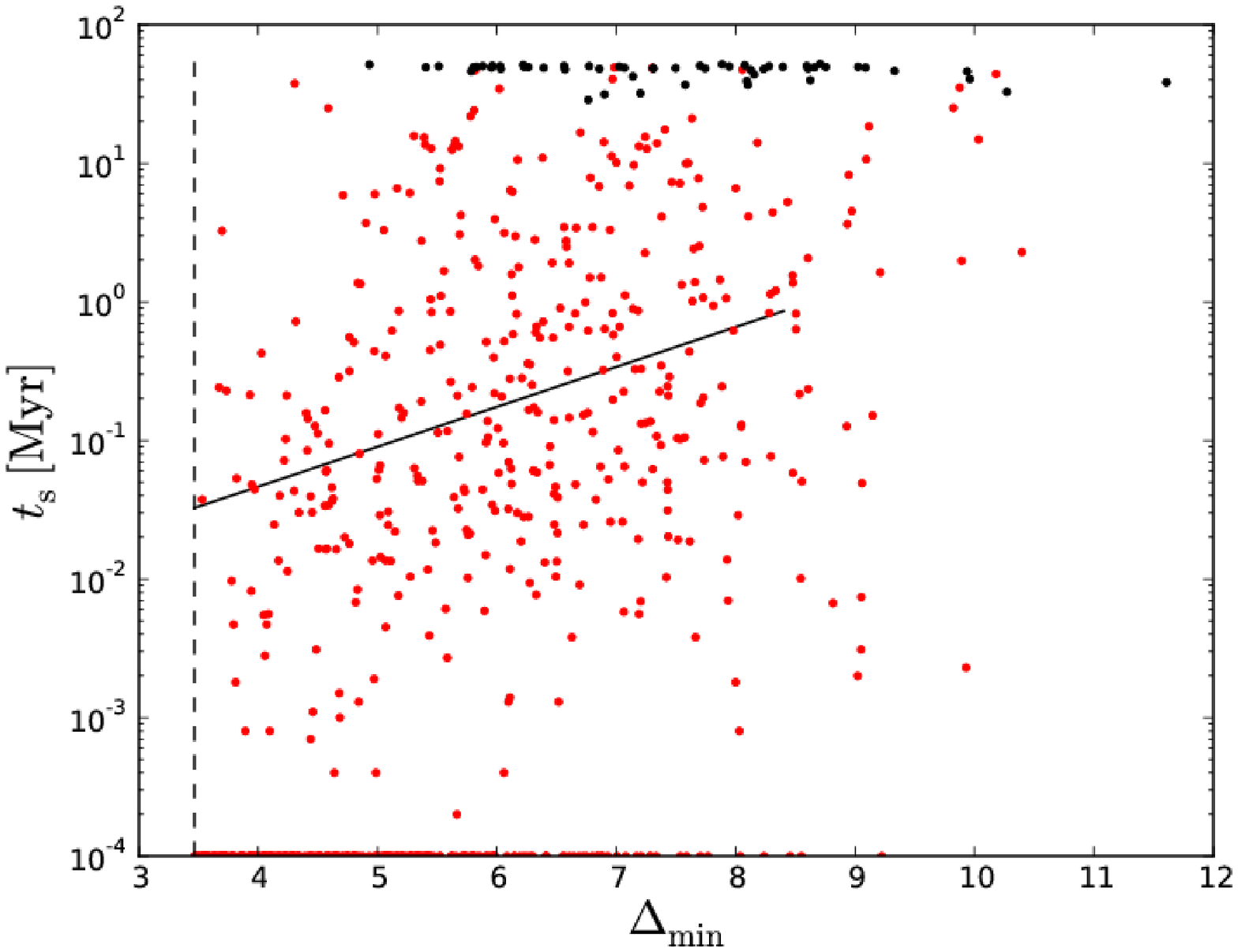}&
\includegraphics[width=8.0cm]{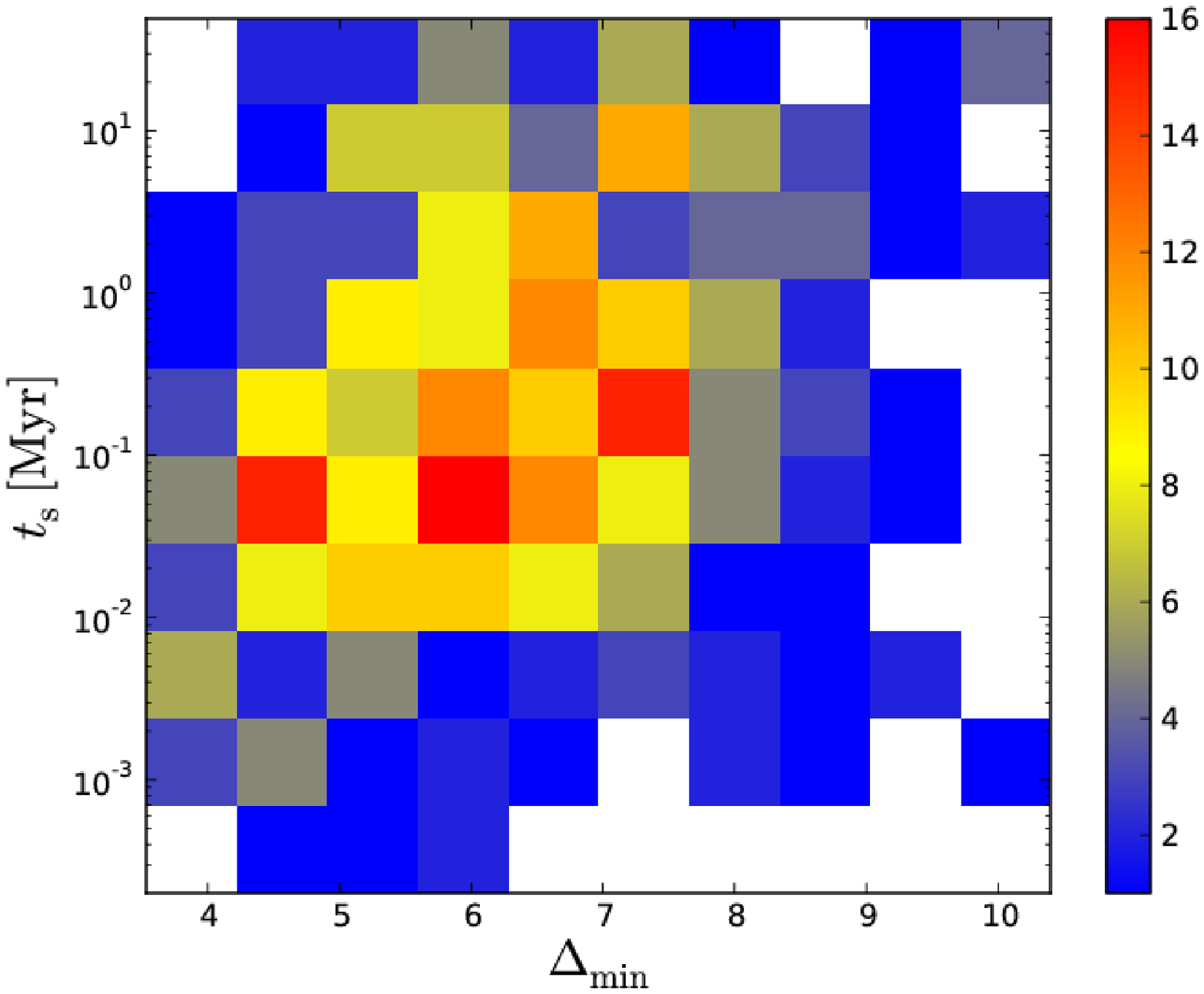}\\
\end{tabular}
\end{center}
\caption{Time to the first instability as a function of minimum dynamical separation between planet pairs, $\Delta_\mathrm{min}$, where the solid line shows the fit described by \eqref{EQ:t_instability_fit} (left).
The red points mark the coordinates of systems with an instability, the black points mark the coordinates of systems that are stable, and the dashed vertical line shows the Hill Stability Criterion, $\Delta>2\sqrt{3}$.
Binned density map of the time to first instability as a function of $\Delta_\mathrm{min}$, excluding systems with $t_\mathrm{s}<10^{-4}$ Myr (right).}
\label{Fig:t_first_instability}
\end{figure}

Figure~\ref{Fig:t_integrations} shows the times to the first instability, defined as a planet-planet collision, and the fraction of systems that are stable at a given integration time.
We perform a least squares fit to the fraction of systems that are stable as a power law,\begin{equation}n_\mathrm{S}(t)=A(t+t_o)^{-\alpha}+C,\label{EQ:PowerLaw}\end{equation} and obtain a best fit with $A=0.22$, $t_o=0.01$, $\alpha=0.19$, and $C=-0.05$.
The histogram shows the distribution of integration times of stable runs, and because not all systems were integrated past the latest time to first instability, we use Bayes' Theorem to project the fraction of systems that are stable at $t>28.7\ \mathrm{Myr}$, the shortest integration time of a stable system (see \S\ref{A:Bayes}).
We find that $96\%$ of the stable systems will remain stable until $t=52\ \mathrm{Myr}$, the longest integration time of a stable system, and $75\%$ will remain stable until $100$ Myr.
Figure~\ref{Fig:t_first_instability} shows the relationship between the initial minimum dynamical separation between planet pairs, $\Delta_\mathrm{min}$, and the time to the first instability, $t_\mathrm{s}$, excluding the systems that were immediately unstable ($t_\mathrm{s}<10^4\ \mathrm{yr}$).
We see that the time to the first instability can be approximated roughly as a log-linear relationship as a function of initial minimum dynamical separation until $\Delta_\mathrm{min}\gtrsim9,$ where half of the dynamical integrations are stable, consistent with \citet{2009Icar..201..381S}.
Fitting \eqref{EQ:t_instability} in the range reported in \citet{2009Icar..201..381S}, $3.4<\Delta<8.4$, we find \begin{equation}\label{EQ:t_instability_fit}\log\left(\frac{t_\mathrm{s}}{1\ \mathrm{Myr}}\right)\simeq0.29\Delta-2.49,\end{equation} although we observe a larger scatter due to constructing our systems with non-uniform masses and dynamical separations, motivating our study for additional predictors of stability.
Finally, we find that no system with initial $\Delta_\mathrm{min}<4.93$ remained stable throughout a dynamical integration.

\subsection{Dependence of Stability on Initial Conditions}
\label{SSec:Stability_Initial_Conditions}
Here we attempt to identify the dependence of stability, defined as the absence of a planet-planet scattering resulting in a collision, on the initial orbital parameters of the system.
Specifically, we perform a $\chi^2$-test of the null hypothesis that the stability of the system is independent of the mass and orbital elements of each planet and various planet-pair characteristics.
The $\chi^2$ value is calculated as \begin{equation}\label{EQ:chi-square}\chi^2=\sum_i\frac{(O_i-E_i)^2}{E_i},\end{equation} where $O_i$ and $E_i$ are the observed and expected number of stable systems in each bin, $i$.
We bin the data so that each bin has roughly the same number of integrations and report the p-values obtained from the $\chi^2$-test, where variables with lower p-values are more likely to influence the stability of the system.
We show the kernel density estimation of the fraction of systems that are stable for the variables with the lowest p-values, smoothing out each observation as a normal distribution with a standard deviation, $\sigma_x(x_i)=|x_i-x_m|$, where $x_m$ is the $m$th closest neighbor to $x_i$, and $m = 2\sqrt{N},$ where $N$ is the total number of runs.
We also show the two and three standard deviation confidence intervals for the null hypothesis, that the stability of the system is independent of the tested variable, which follows a binomial distribution, $\sigma_\mathrm{S}(x)=\sqrt{P(1-P)/n(x)}$, where $P=0.078$ is the fraction of systems that are stable and $n(x)$ is the kernel density estimation of the number of integrations at $x$.

\begin{figure}[htp]
\begin{center}
\begin{tabular}{cc}
\includegraphics[width=8.0cm]{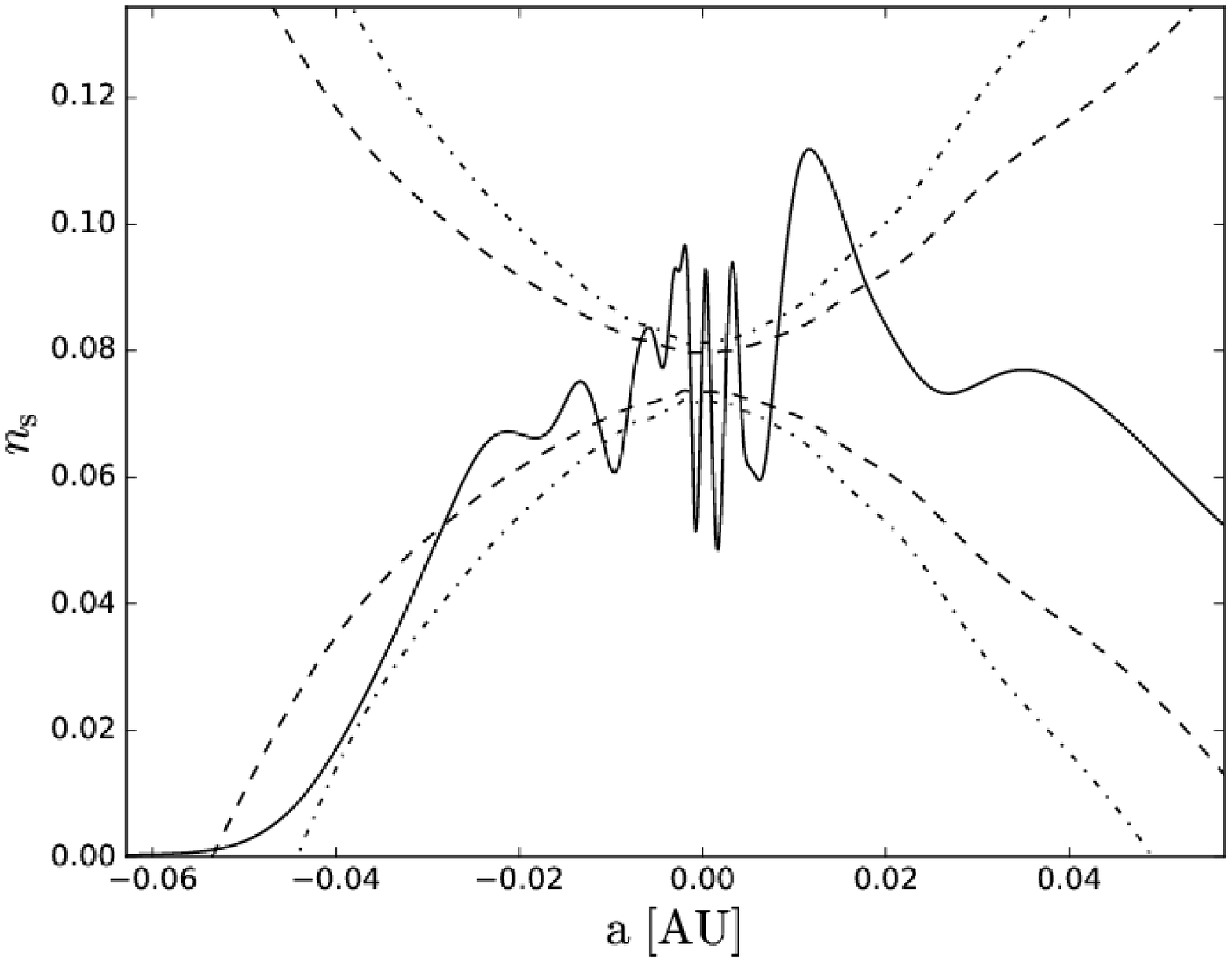}&
\includegraphics[width=8.0cm]{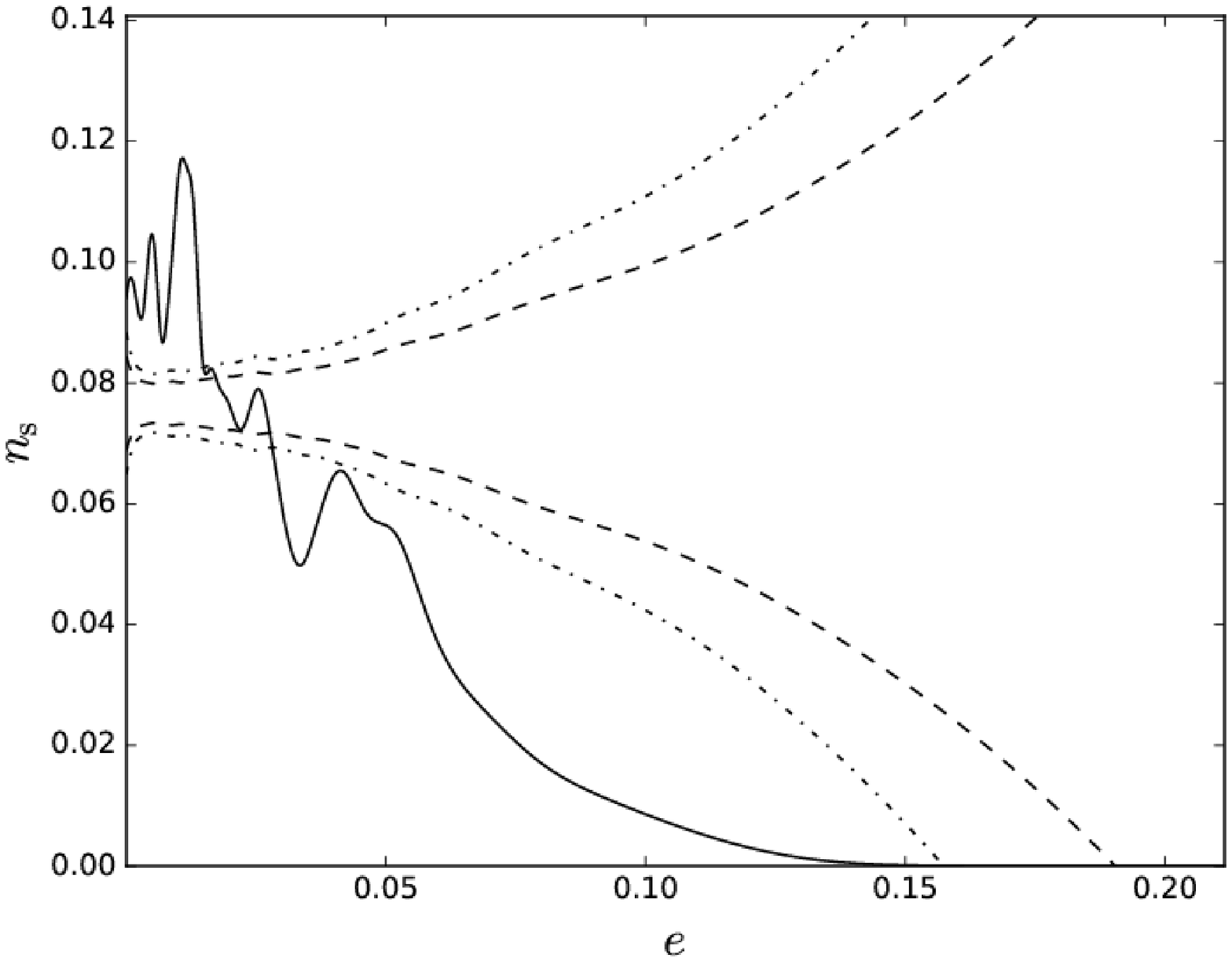}\\
\\
\includegraphics[width=8.0cm]{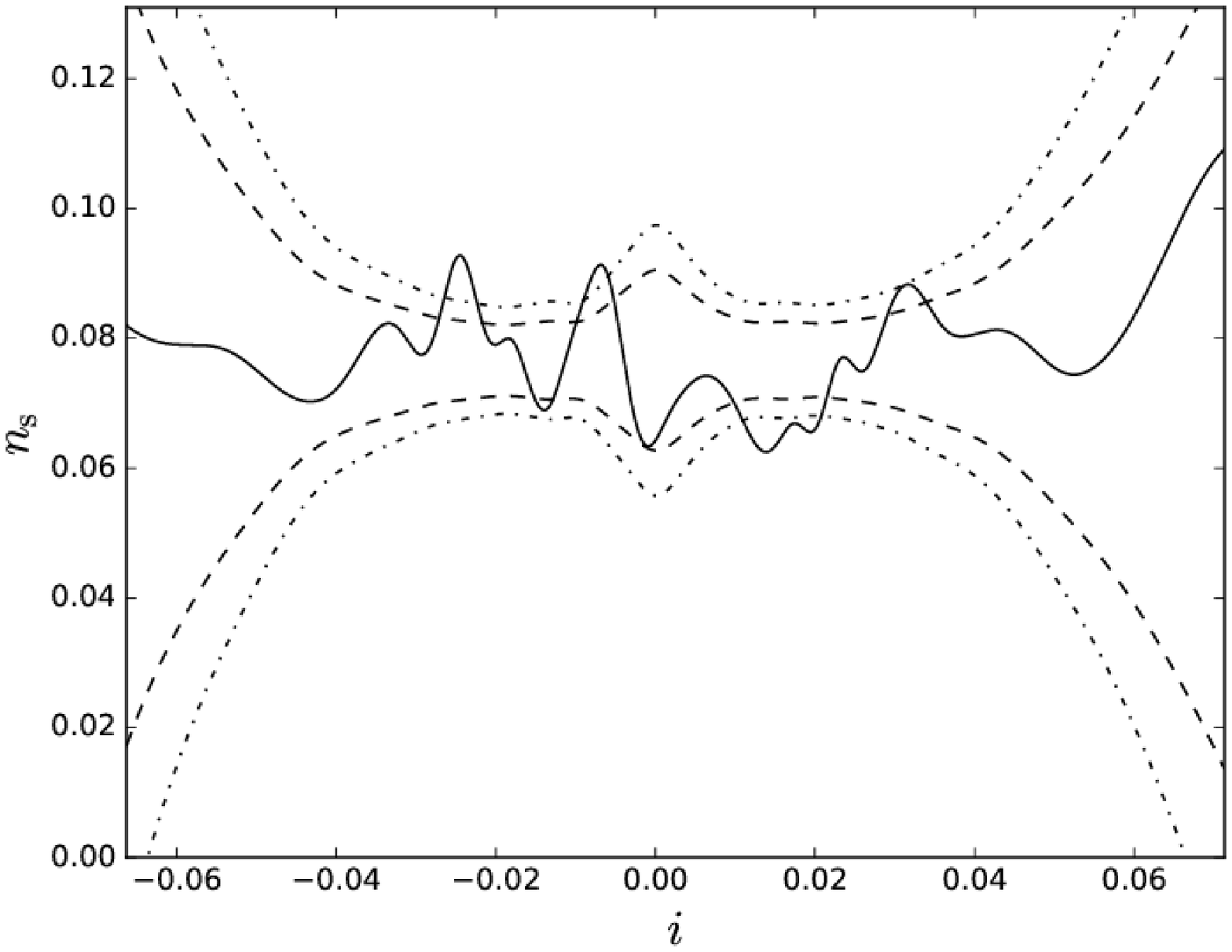}&
\includegraphics[width=8.0cm]{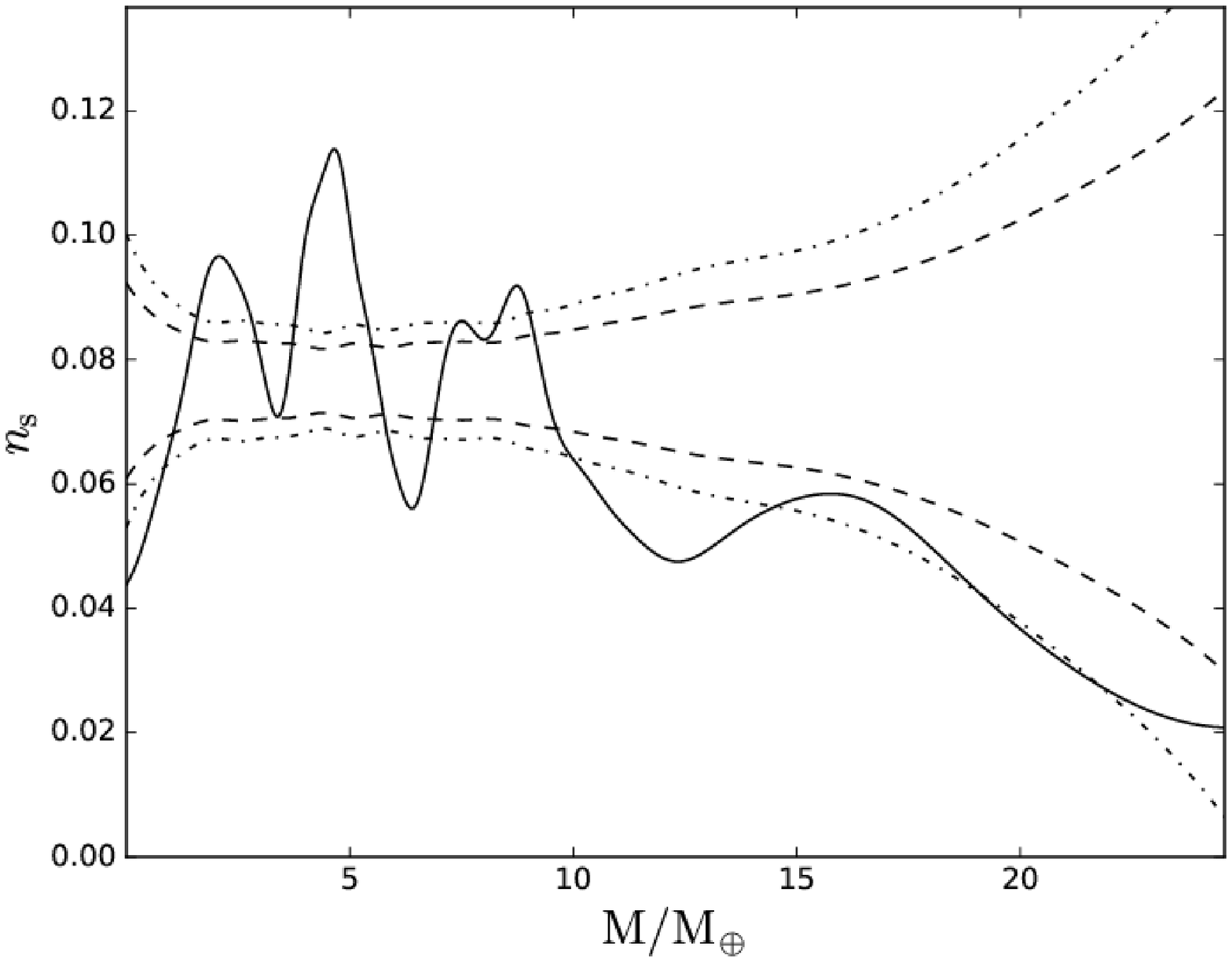}\\
\end{tabular}
\end{center}
\caption{The solid lines show the kernel density estimation of the fraction of systems that are stable as a function of the semi-major axes (top left; shown as the difference from the nominal values reported in \citealt{2011Natur.470...53L}), the eccentricities (top right), the inclinations (bottom left), and the masses (bottom right) of the planets, excluding Kepler-11g.
Also shown are the two (dashed lines) and three (dash-dotted lines) standard deviation confidence intervals assuming the null hypothesis, that the stability of the system is independent of the tested variable, is true.
The width of the confidence intervals scales inversely with the sample density, $\sigma_\mathrm{S}(x)\propto 1/\sqrt{n(x)}$, where a two standard deviation confidence interval width of $1\%$ corresponds to $n(x)\sim700$.}
\label{Fig:Planet_KDE}
\end{figure}

Table~\ref{TBL:Chi-2_planets} shows the p-values from the $\chi^2$-test of the null hypothesis that the stability of the system is independent of the semi-major axis, eccentricity, longitude of ascending node, argument of periapsis, and mass of each planet in the system and all the planets, excluding Kepler-11g.
Figure~\ref{Fig:Planet_KDE} shows the kernel density estimation of the fraction of systems that are stable for the semi-major axis, eccentricity, inclination, and masses of all the planets, excluding Kepler-11g, and the sample density, which is inversely proportional to the width of the shown variance.
We find that, in general the stability of the system is least likely to be independent of the semi-major axes and especially the eccentricities of the planets, with the maximum stability occurring at non-zero values of eccentricity due to a lower libration amplitude.
The stability is comparatively less sensitive to the planets' masses and inclinations, consistent with \citet{2015ApJ...807...44P}.

\begin{deluxetable}{ccccccccc}
\tablewidth{10cm}
\tabletypesize{\footnotesize}
\tablecolumns{8}
\tablecaption{p-values from $\chi^2$-Test of Independence of Stability from Planet Mass and Orbital Parameters \label{TBL:Chi-2_planets}}
\tablehead{
    \colhead{$Planet$} & \colhead{$a$} & \colhead{$e$} & \colhead{$i$} & \colhead{$\Omega$} & \colhead{$\omega$} & \colhead{$M$} & \colhead{$m$}}
\startdata
$ Kepler-11b $ & $ 0.079 $ & $ 0.009 $ & $ 0.876 $ & $ 0.831 $ & $ 0.733 $ & $ 0.172 $ & $ 0.645 $ & \\
$ Kepler-11c $ & $ 0.238 $ & $ 0.267 $ & $ 0.978 $ & $ 0.764 $ & $ 0.945 $ & $ 0.680 $ & $ 0.117 $ & \\
$ Kepler-11d $ & $ 0.299 $ & $ 0.003 $ & $ 0.497 $ & $ 0.747 $ & $ 0.687 $ & $ 0.365 $ & $ 0.238 $ & \\
$ Kepler-11e $ & $ 0.026 $ & $ 0.016 $ & $ 0.483 $ & $ 0.538 $ & $ 0.245 $ & $ 0.303 $ & $ 0.240 $ & \\
$ Kepler-11f $ & $ 0.126 $ & $ 0.090 $ & $ 0.650 $ & $ 0.331 $ & $ 0.242 $ & $ 0.008 $ & $ 0.296 $ & \\
$ Kepler-11g $ & $ 0.483 $ & $ 0.518 $ & $ 0.220 $ & $ 0.518 $ & $ 0.993 $ & $ 0.747 $ & $ 0.098 $ & \\
\midrule
$ All\ Planets^2 $ & $ 0.107 $ & $ 0.024 $ & $ 0.999 $ & $ 0.454 $ & $ 0.980 $ & $ 0.731 $ & $ 0.352 $ & \\
\enddata

\tablenotetext{1}{p-values from the $\chi^2$-test of independence of stability from the mass and orbital parameters of the planets, where $a$ is the semi-major axis, $e$ is the eccentricity, $i$ is the inclination, $\Omega$ is the longitude of ascending node, $\omega$ is the argument of periastron, and $m$ is the mass of the planet.}
\tablenotetext{2}{The aggregate values omit Kepler-11g from the calculations, as this planet is much less dynamically active than the inner 5 planets. The summary value for $a$ is calculated using the difference between the semi-major axis used in the calculation compared to the nominal values reported in \citet{2011Natur.470...53L}.}
\end{deluxetable}

\begin{figure}[htp]
\begin{center}
\begin{tabular}{cc}
\includegraphics[width=8.0cm]{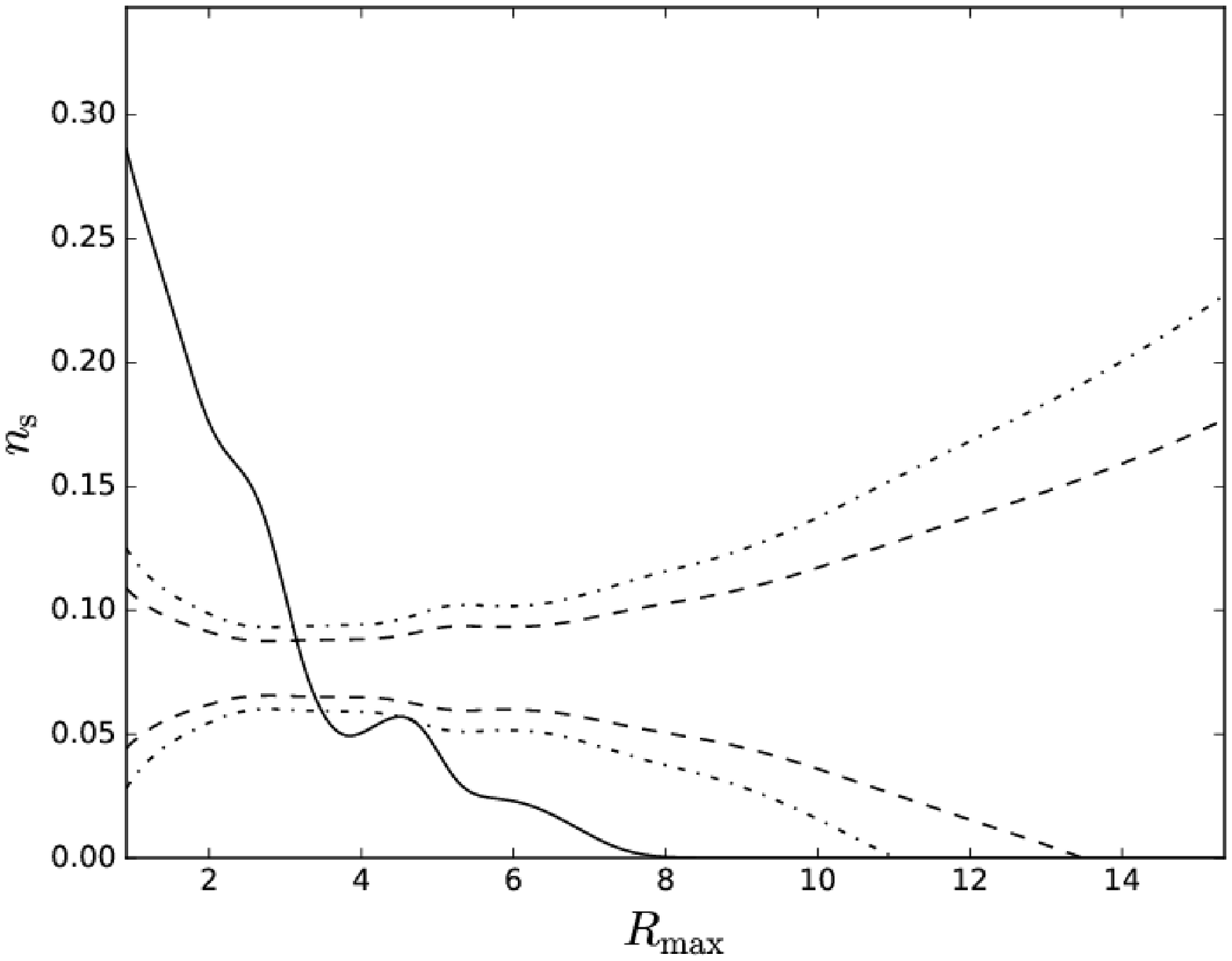}&
\includegraphics[width=8.0cm]{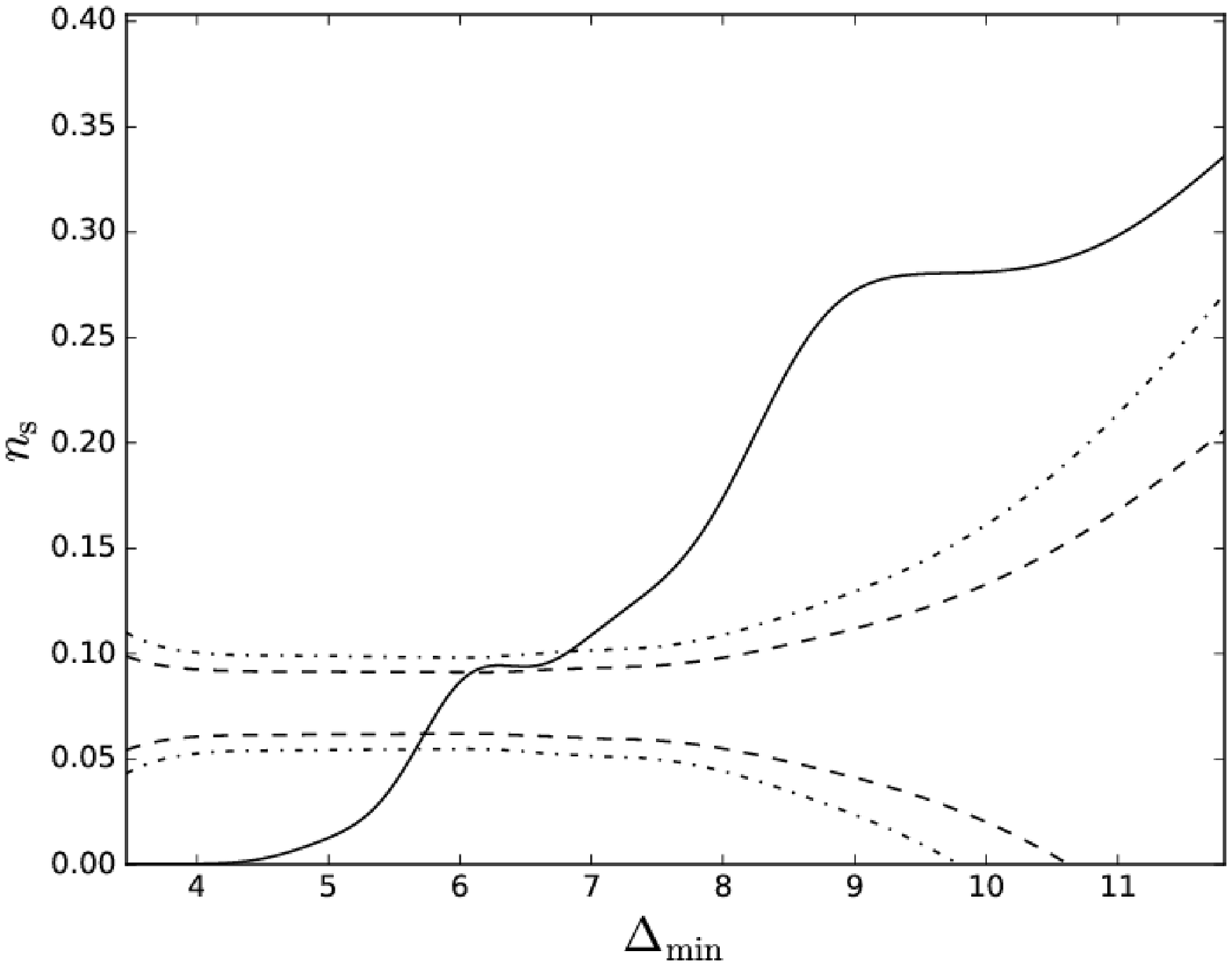}\\
\\
\includegraphics[width=8.0cm]{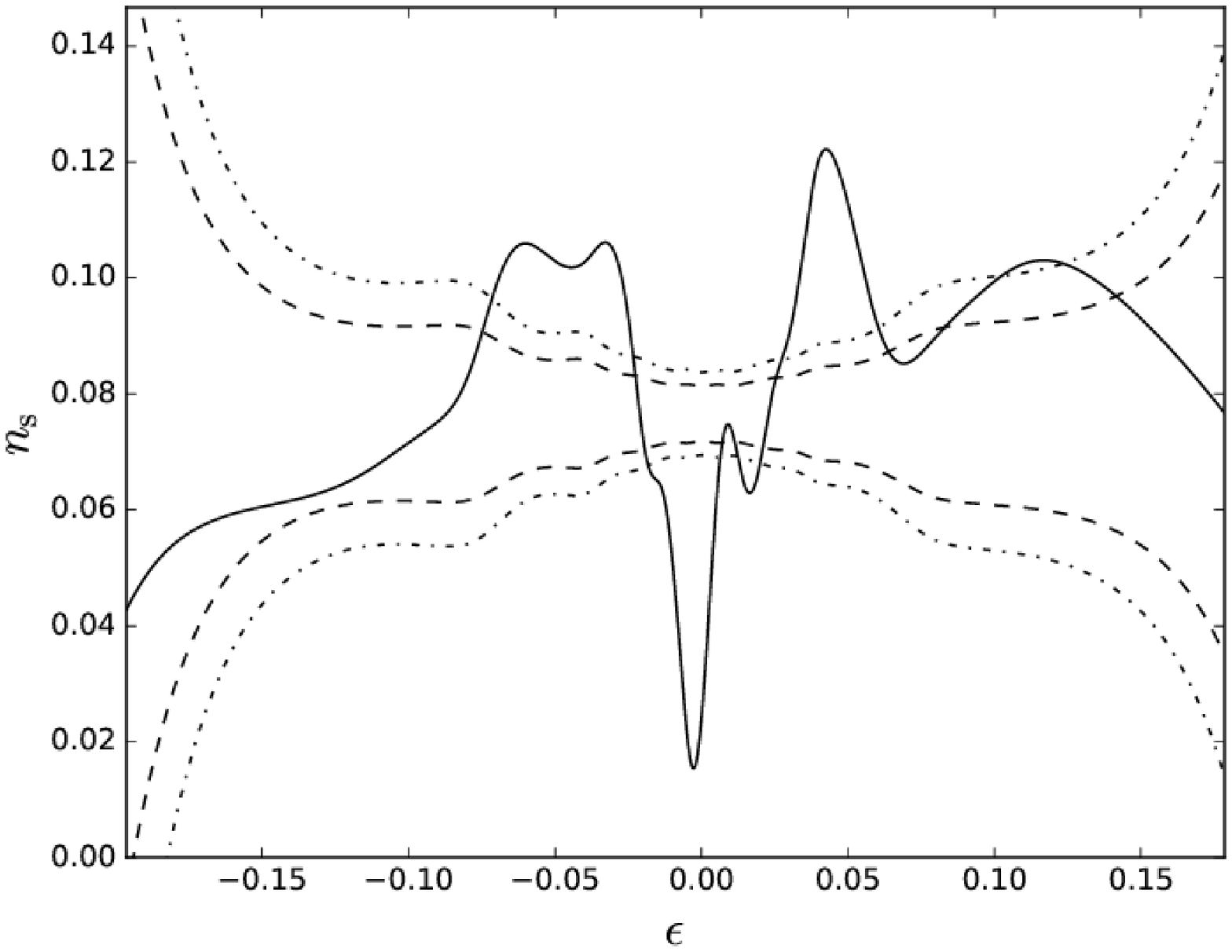}&
\includegraphics[width=8.0cm]{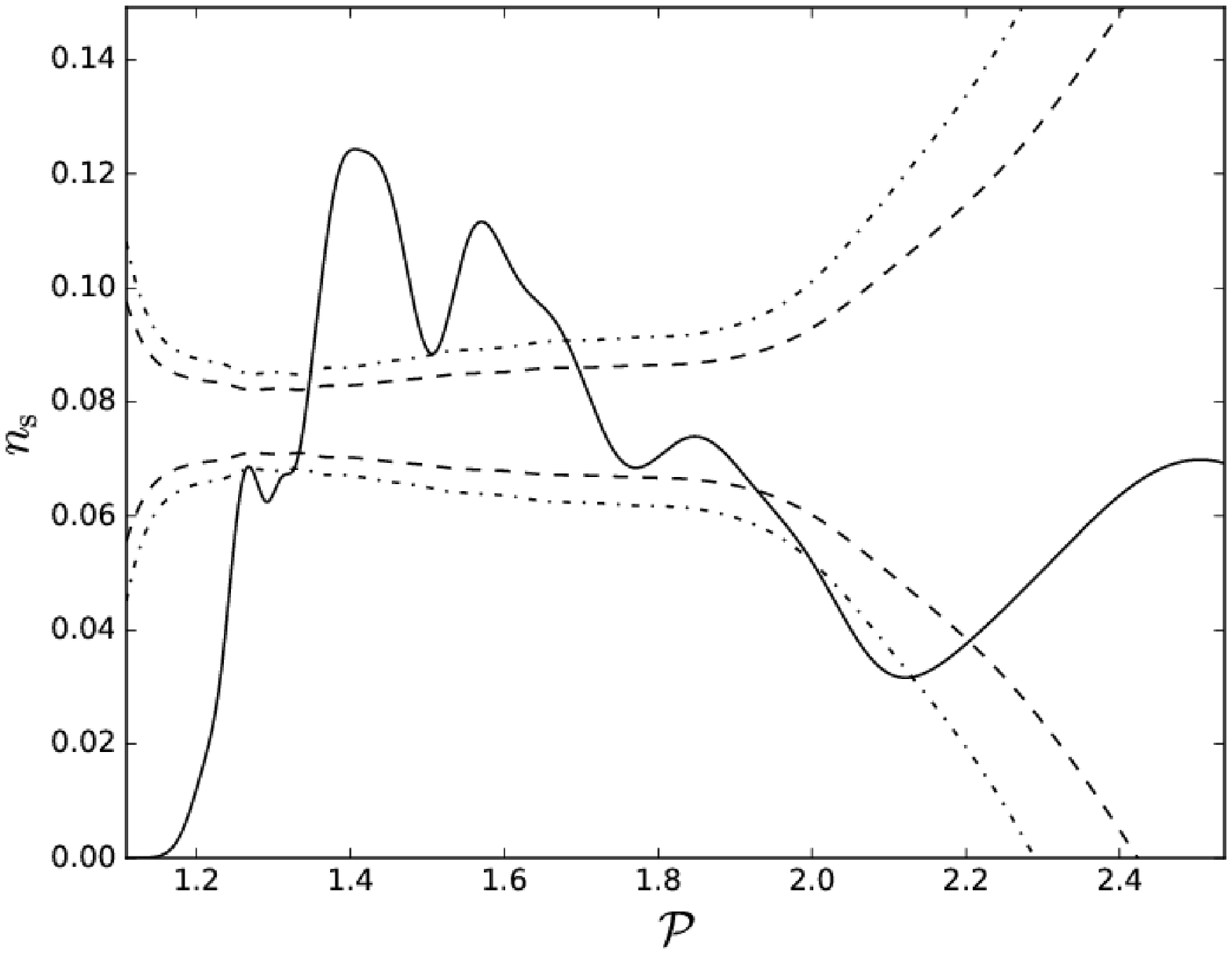}\\
\end{tabular}
\end{center}
\caption{The solid lines show the kernel density estimation of the fraction of systems that are stable as a function of the maximum magnitude of the action-angle variables (top left), the minimum dynamical separation (top right), the distance to MMR (bottom left), and the period ratio (bottom right) of the planet-pairs, excluding Kepler-11fg.
Also shown are the two (dashed lines) and three (dash-dotted lines) standard deviation confidence intervals assuming the null hypothesis, that the stability of the system is independent of the tested variable, is true.
The width of the confidence intervals scales inversely with the sample density, $\sigma_\mathrm{S}(x)\propto 1/\sqrt{n(x)}$, where a two standard deviation confidence interval width of $1\%$ corresponds to $n(x)\sim700$.}
\label{Fig:PlanetPair_KDE}
\end{figure}

Table~\ref{TBL:Chi-2_pairs} shows the p-values from the $\chi^2$-test of the null hypothesis that stability of the system is independent of the magnitude of the action-angle variables, dynamical separation, distance to mean motion resonance (MMR), and period ratio of each planet pair and the maximum action-angle variable, minimum dynamical separation, and minimum absolute distance from MMR in the system.
We find that, in general the stability of the system is less likely to be independent of the planet-pair characteristics than the individual planets' masses and orbital parameters.
We define the magnitude of the action-angle variables as \begin{equation}\label{EQ:R}R=\sqrt{X^2+Y^2},\end{equation} where the canonical action-angle variables, $X$ and $Y$, are defined from reducing the original Hamiltonian \begin{equation}H\simeq\left(\frac{p_1^2}{2m_1}-\frac{GM_*m_1}{|\mathbf{r_1}|}\right)+\left(\frac{p_2^2}{2m_2}-\frac{GM_*m_2}{|\mathbf{r_2}|}\right)-Gm_1m_2\left(\frac{1}{|\mathbf{r_1}-\mathbf{r_2}|}-\frac{\mathbf{r_2}\cdot\mathbf{r_1}}{|\mathbf{r_2}|^3}\right)\end{equation} to \begin{equation}H'=-\frac{1}{2}\left[\frac{1}{2}(X^2+Y^2)-\Gamma'\right]^2-X,\end{equation} where $p_i$ is the momentum, $m_i$ is the mass, and $\mathbf{r_i}$ is the position coordinate of the planet, where the subscript $1$ designates the inner planet and the subscript $2$ designates the outer planet, $M_*$ is the mass of the star, set at the origin, $G$ is the gravitational constant, and $\Gamma'$ is determined by the distance from resonance (see \S2 in \citealt{2013ApJ...774..129D} for more details).
The distance to MMR is defined as \begin{equation}\epsilon=min\left(\mathcal{P}\frac{j}{j+1}-1,\mathcal{P}\frac{1}{j}\right),\label{EQ:Epsilon}\end{equation} where $j\in\mathbb{Z}$ is the resonance index or integer period ratio and $\mathcal{P}=P_2/P_1$, where $P_1$ and $P_2$ are the orbital periods of the inner and outer planets respectively \citep{2012ApJ...756L..11L}.
Figure~\ref{Fig:PlanetPair_KDE} shows the kernel density estimation of the fraction of systems that are stable for the planet-pair characteristics described in Table~\ref{TBL:Chi-2_pairs} and the sample density, which is inversely proportional to the width of the shown variance.
We find that $R_\mathrm{max}$ and $\Delta_\mathrm{min}$ are the strongest predictors of stability, with the largest range of responses: systems are generally more stable at lower values of $R_\mathrm{max}$, with no stable systems with $R>6.17$, and at higher values of $\Delta_\mathrm{min}$, with no stable systems with $\Delta<4.93$.
For the inner planet-pairs (excluding Kepler-11fg), $R$ is a more consistent predictor of stability than $\Delta$, the metric most commonly used to characterize the stability of planetary systems.
We also find a significant trough in stability at $\epsilon = 0$ with peaks just inside and outside of the 3:2 period ratio.

\begin{deluxetable}{cccccc}
\tablewidth{7.5cm}
\tabletypesize{\footnotesize}
\tablecolumns{5}
\tablecaption{p-values from $\chi^2$-Test of Independence of Stability from Planet-Pair Characteristics \label{TBL:Chi-2_pairs}}
\tablehead{
    \colhead{$Planet\ Pair$} & \colhead{$R$} & \colhead{$\Delta$} & \colhead{$\epsilon$} & \colhead{$\mathcal{P}$}}
\startdata
$ Kepler-11bc $ & $ 0.029 $ & $ 0.006 $ & $ 0.010 $ & $ 0.014 $ & \\
$ Kepler-11cd $ & $ 0.010 $ & $ 0.530 $ & $ 0.204 $ & $ 0.269 $ & \\
$ Kepler-11de $ & $ 1\times10^{-5} $ & $ 0.003 $ & $ 0.059 $ & $ 0.029 $ & \\
$ Kepler-11ef $ & $ 0.090 $ & $ 0.443 $ & $ 0.245 $ & $ 0.103 $ & \\
$ Kepler-11fg $ & $ 0.560 $ & $ 0.269 $ & $ 0.443 $ & $ 0.556 $ & \\
\midrule
$ All\ Pairs^2 $ & $ 2\times10^{-5} $ & $ 0.008 $ & $ 0.002 $ & $ 3\times10^{-4} $ & \\
\midrule
$ $ & $ R_\mathrm{max} $ & $ \Delta_\mathrm{min} $ & $ |\epsilon|_\mathrm{min} $ & \\
\midrule
$ System $ & $ 7\times10^{-8} $ & $ 5\times10^{-8} $ & $ 0.040$ & \\
\enddata
\tablenotetext{1}{p-values from a $\chi^2$-test of independence of stability from various planet-pair characteristics, where $R$ is the magnitude of the action-angle variables, defined in \eqref{EQ:R}, $\Delta$ is the dynamical separation, defined in \eqref{EQ:DynamicalSeparation}, $\epsilon$ is the distance to MMR, defined in \eqref{EQ:Epsilon}, and $\mathcal{P}$ is the period ratio. $R_\mathrm{max}$, $\Delta_\mathrm{min}$, and $|\epsilon|_\mathrm{min}$ are the minimum and maximum values among planet-pairs in the system.}
\tablenotetext{2}{The aggregate values omit the Kepler-11fg planet-pair from the calculations, as this planet pair is much less dynamically active than the inner 5 planets.}
\end{deluxetable}

\subsection{Architecture}
\label{SSec:Architecture}
\begin{figure}[htp]
\begin{center}
\begin{tabular}{cc}
\includegraphics[width=8.0cm]{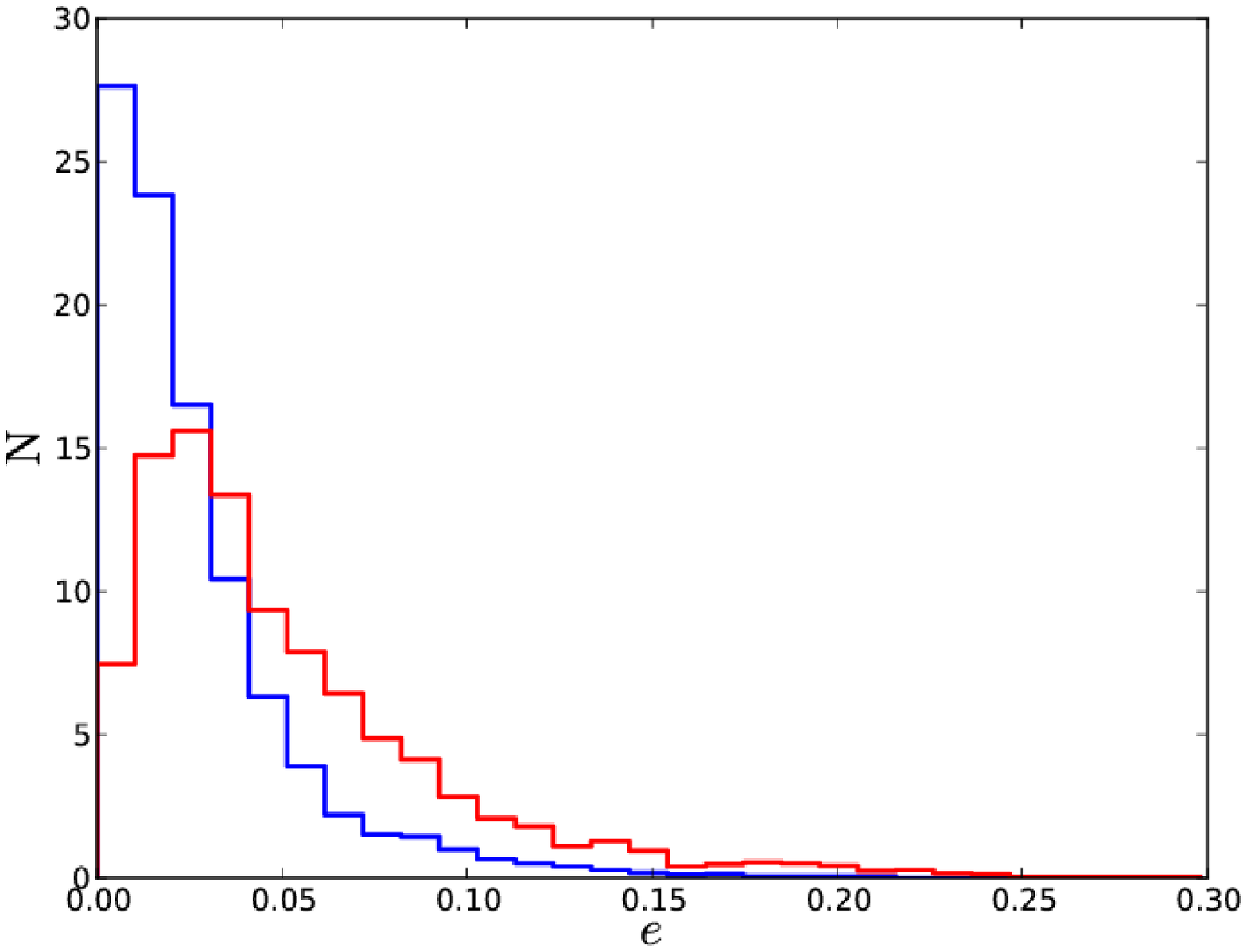}&
\includegraphics[width=8.0cm]{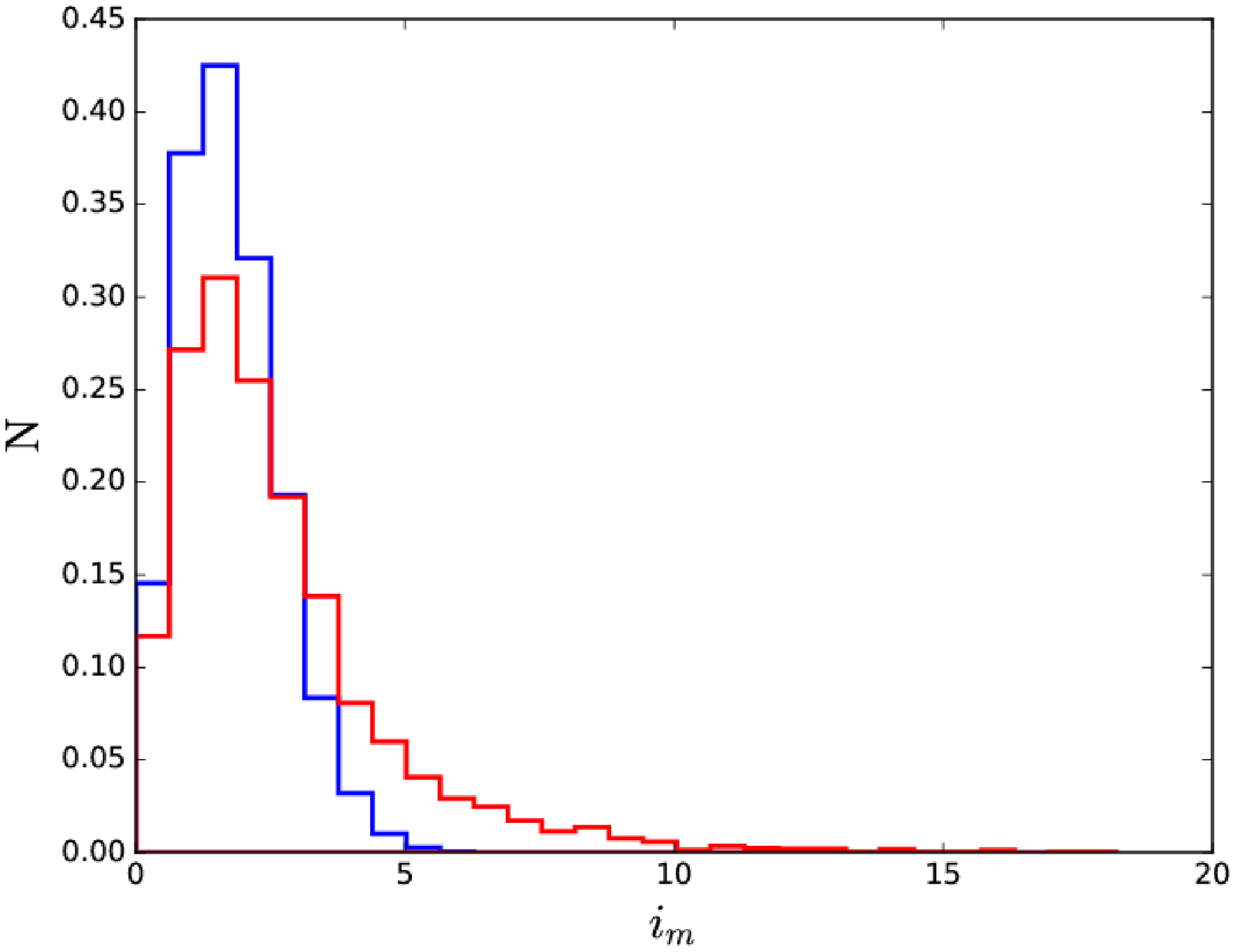}\\
\end{tabular}
\end{center}
\caption{Initial (blue) and final (red) distributions of eccentricities (left) and mutual inclinations (right) of systems with an instability.
We see that, in general, an instability tends to increase the eccentricities and mutual inclinations within the system.}
\label{Fig:e_i_histogram}
\end{figure}

Here we examine in more detail the architecture of the remnant systems.
Specifically, we compare the initial and final eccentricity and mutual inclination distributions for systems with an instability and analyze how the dynamical separation evolves and compares to the currently observed Kepler sample.
Figure~\ref{Fig:e_i_histogram} shows the initial and final distributions of both the mutual inclination and eccentricity for systems with an instability.
We aggregate the mutual inclinations between each combination of planet pairs in the system, determined from the reported inclination, $i$, and longitude of ascending node, $\Omega$,
 \begin{equation}i_\mathrm{m}=\cos^{-1}\left(\frac{\mathbf{u}_j\cdot\mathbf{u}_k}{|\mathbf{u}_j||\mathbf{u}_k|}\right),\end{equation} where $i_\mathrm{m}$ is the mutual inclination between planets $j$ and $k$, and \begin{equation}\mathbf{u}=\begin{pmatrix}-\sin(\Omega)\sin(i)\\\cos(\Omega)\cos(i)\\\cos(i)\end{pmatrix}.\end{equation}
We find that the remaining planets in the unstable realizations have larger eccentricities and inclinations as a result of the planet-planet interactions, and the presence of excited orbits are a potential indicator of previous planet-planet instabilities.

\begin{figure}[htp]
\begin{center}
\begin{tabular}{cc}
\includegraphics[width=8.0cm]{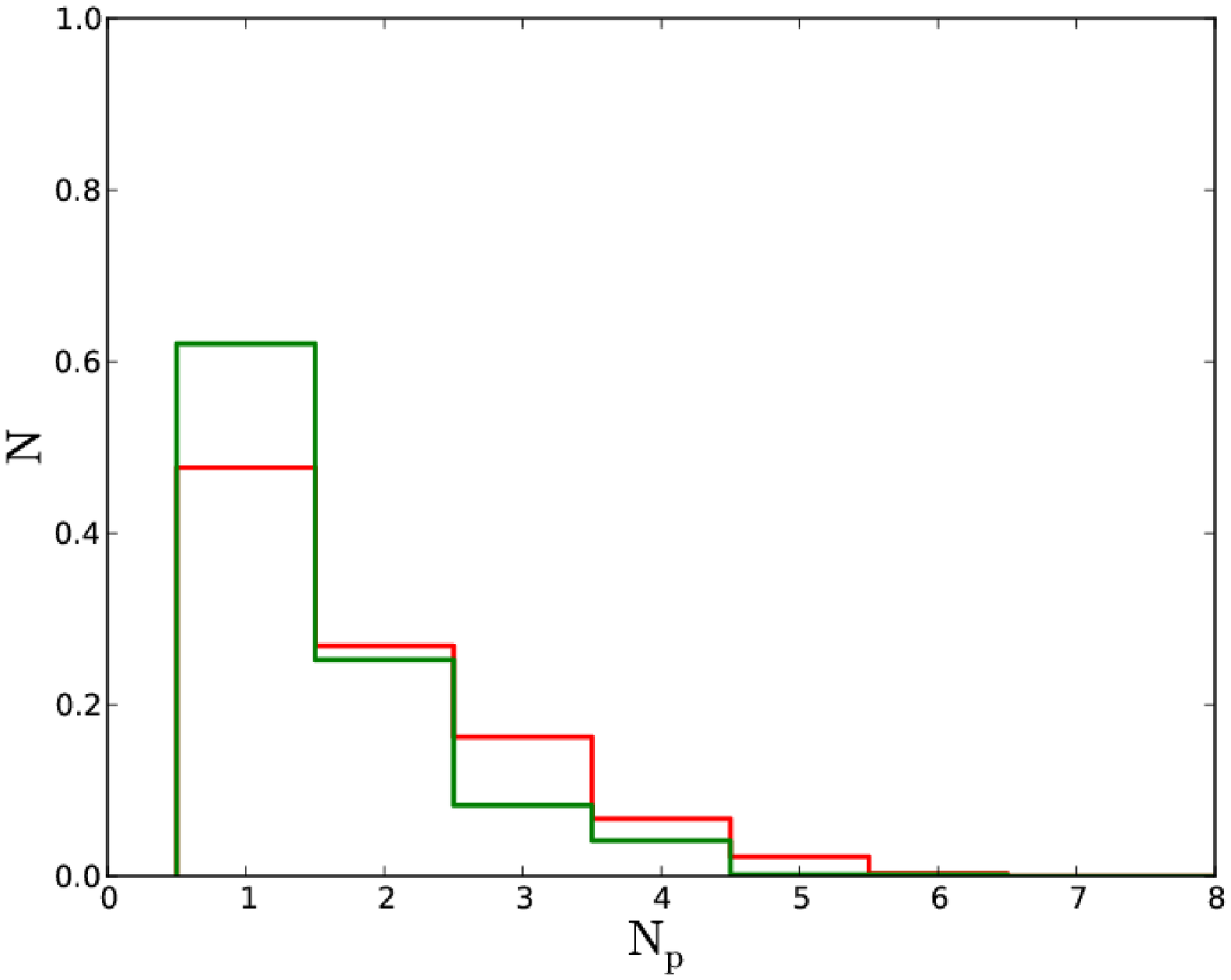}
\includegraphics[width=8.0cm]{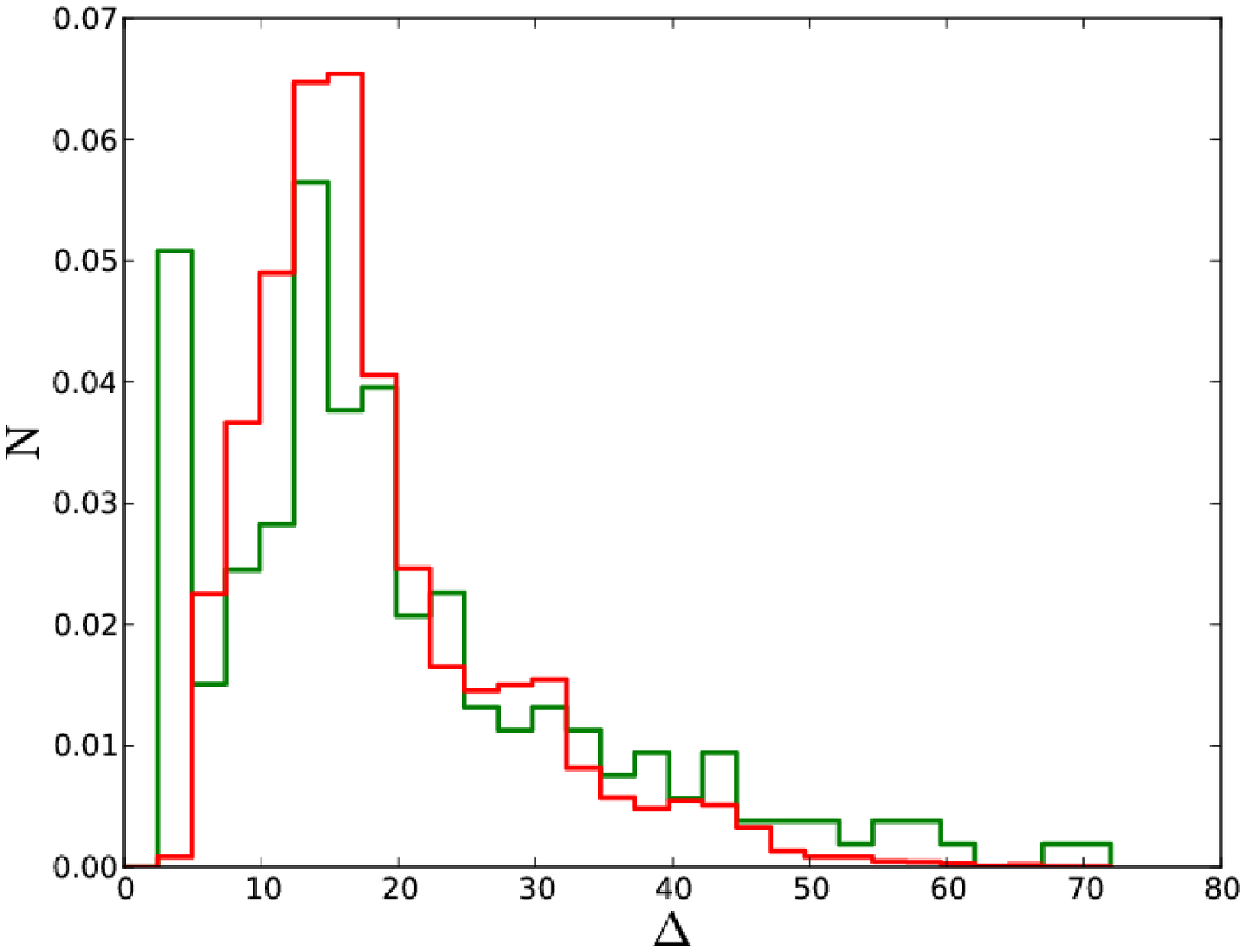}\\
\end{tabular}
\end{center}
\caption{
Observed distributions of planet multiplicity (left) and dynamical separation (right) from the Kepler sample (green) and all dynamical integration remnants (red).
The observed distributions from the dynamical integration remnants are created by randomly choosing lines of sight to determine which planets transit (see \S\ref{A:Transit}).
Both distributions from the dynamical integration remnants agree reasonably well with the distributions from the observed Kepler sample, and the deficit of single-planet transits and low dynamical separations may suggest a separate population of very compact, low-multiplicity systems}
\label{Fig:Sep_Observed}
\end{figure}

We generate a synthetic population of $\sim2\times10^5$ observed systems from our dynamical integrations, sampling each integration 300 times, by randomly picking lines of sight to the remnant systems and determining which planets transit.
We determine that a planet transits if \begin{equation}r\sin(d/r)\le R_*,\end{equation} where $r$ is the orbital separation of the planet, $d$ is the smallest distance from the line of sight to the orbit, and $R_*=R_\odot$ is the radius of the star (for additional details, see \S\ref{A:Transit}).
Table~\ref{TBL:Transit_Stats} details the rates of the observed multiplicities and how often each planet transits.
We find that $\sim6\%$ of systems have transiting planets, and of the systems with a transit, $48\%$ have a single transiting planet, $27\%$ have two transiting planets, $16\%$ have three transiting planets, $7\%$ have four transiting planets, $2\%$ have five transiting planets, and $0.3\%$ have six transiting planets.
Figure~\ref{Fig:Sep_Observed} shows a comparison of the distributions of planet multiplicity and dynamical separation between the synthetically generated observed population and the observed Kepler sample.
We see that both the observed distributions of planet multiplicity and dynamical separation of the synthetically-generated systems agree reasonably well with the observed distributions from the Kepler sample.
We see an expected deficit of single-planet systems, a small deficit at large $\Delta$, and a large deficit at small $\Delta$, which may be due to the high multiplicity of our initial conditions; low-multiplicity systems are more likely to remain stable at lower dynamical separations than high-multiplicity systems.
These deficits may suggest the presence of a separate population of very compact, low-multiplicity systems \citep{2014ApJ...793L...3V}.
These results are consistent with our hypothesis that a subset of the currently observed Kepler systems could be remnants of even more tightly-packed systems that underwent instabilities.

\begin{deluxetable}{ccc|ccc}
\tablewidth{10.0cm}
\tabletypesize{\footnotesize}
\tablecolumns{6}
\tablecaption{Transit Statistics from Synthetic Population\label{TBL:Transit_Stats}}
\tablehead{
    \colhead{$Multiplicity$} & \colhead{$\% Observed$} & \colhead{$\% Underlying$} & \colhead{$Planet$} & \colhead{$\% Observed$}}
\startdata
$ 1 $ & $ 47.6\% $ & $  0.00\% $ & $ Kepler-11b $ & $ 36.7\% $ & \\
$ 2 $ & $ 26.8\% $ & $  1.01\% $ & $ Kepler-11c $ & $ 67.6\% $ & \\
$ 3 $ & $ 16.3\% $ & $ 28.39\% $ & $ Kepler-11d $ & $ 23.8\% $ & \\
$ 4 $ & $  6.7\% $ & $ 44.09\% $ & $ Kepler-11e $ & $ 33.7\% $ & \\
$ 5 $ & $  2.3\% $ & $ 18.73\% $ & $ Kepler-11f $ & $ 11.9\% $ & \\
$ 6 $ & $  0.3\% $ & $  7.78\% $ & $ Kepler-11g $ & $ 16.3\% $ & \\
\enddata
\tablenotetext{1}{Distribution of planet multiplicities seen in the synthetically generated population of observed systems compared to the underlying population, and distribution of the transiting planets.
The number of transits per planet is affected by mergers, in which the lower-mass planet is removed from the integration.}
\end{deluxetable}

\section{Collisions}
\label{Sec:Collisions}

\subsection{Characterization}
\label{SSec:Characterization}
Of the 694 simulated systems, 640 exhibit at least one planet-planet interaction, resulting in a total of 1361 planet-planet collisions.
We are primarily interested in the details of the first planet-planet collision of each system to generate initial conditions for 3-D hydrodynamic calculations, two of which are shown in \S\ref{SSec:SPH_Calculation}.
We find that a majority of first collisions occur between neighboring planets; specifically Kepler-11b and Kepler-11c ($27\%$), Kepler-11c and Kepler-11d ($26\%$), and Kepler-11d and Kepler-11e ($22\%$).
For each close encounter resulting in a collision we record the distance of closest approach, collision energy, and position and velocity coordinates.
A collision occurs in the integration when the planets have a minimum separation, $b$, less than the sum of the physical radii and is treated with the sticky-sphere approximation (see \citealt{1996Icar..119..261C}), resulting in a single surviving planet with mass equal to the sum of the two colliding planets' masses,
\begin{equation}m_\mathrm{f}=m_1+m_2,\end{equation} energy lost, \begin{equation}\Delta E=\frac{1}{2}\frac{m_1m_2}{m_1+m_2}|\mathbf{v}_\mathrm{1,i}-\mathbf{v}_\mathrm{2,i}|^2 - \frac{Gm_1m_2}{b},\end{equation} final spin angular momentum,
\begin{equation}\mathbf{s}_\mathrm{f}=\mathbf{s}_\mathrm{i,1}+\mathbf{s}_\mathrm{i,2}+\mathbf{r}\times\mathbf{p},\end{equation}
 final position and velocity coordinates determined by conserving center of mass and momentum, \begin{equation}\mathbf{x}_\mathrm{f}=\mathbf{x}_\mathrm{1,i}\frac{m_1}{m_1+m_2}+\mathbf{x}_\mathrm{2,i}\frac{m_2}{m_1+m_2},\end{equation} \begin{equation}\mathbf{v}_\mathrm{f}=\mathbf{v}_\mathrm{1,i}\frac{m_1}{m_1+m_2}+\mathbf{v}_\mathrm{2,i}\frac{m_2}{m_1+m_2},\end{equation} and density of the higher-mass planet. In all cases, the index given to a merger product is the index of the more massive of the two colliding parents.

\begin{figure}[htp]
\begin{center}
\begin{tabular}{cc}
\includegraphics[width=16.0cm]{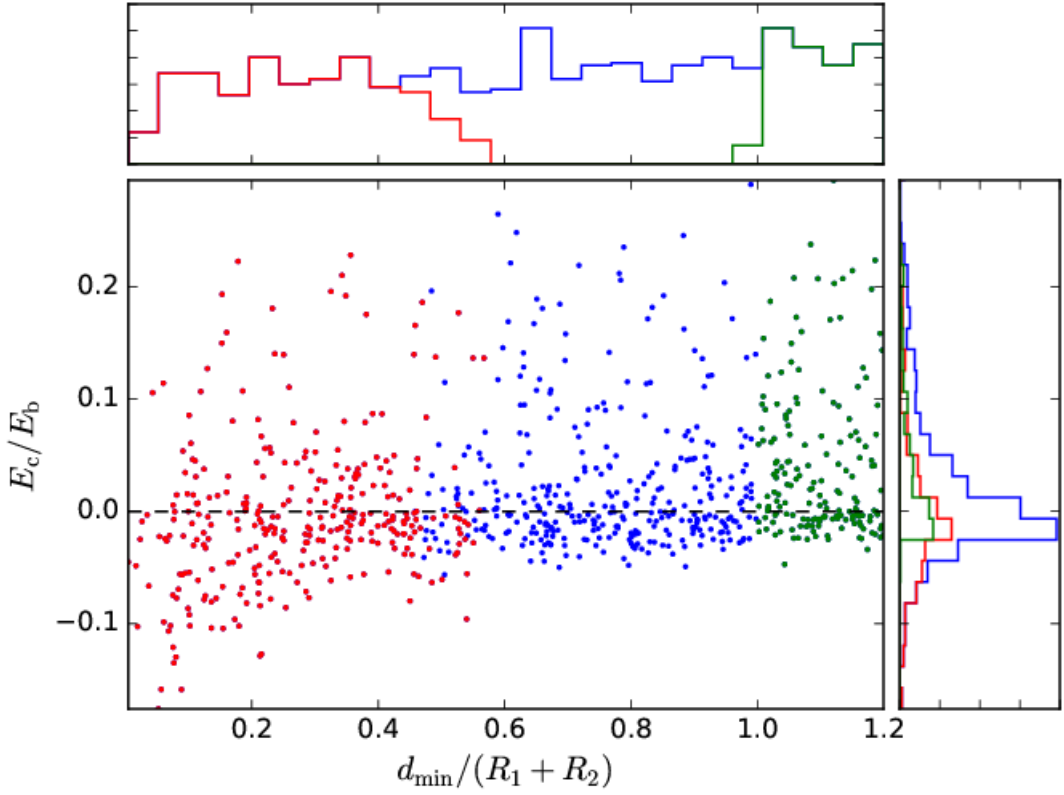}
\end{tabular}
\end{center}
\caption{Scatter plot showing the distance of closest approach, $d_\mathrm{min}$, in units of the sum of the planets' physical radii, and collision energy, $E_\mathrm{c}=E_\mathrm{k}+E_\mathrm{g}$ in units of the binding energy of both planets up to a coefficient, $E_\mathrm{b}=Gm_1^2/R_1+Gm_2^2/R_2$, where $E_\mathrm{k}$ and $E_\mathrm{g}$ are the kinetic and gravitational potential energy of the planets in the center of mass frame, ignoring the host star, for the first collision or near miss in each integration. The histogram above the scatter-plot shows the distribution of distances to closest approach and the histogram to the right of the scatter-plot shows the distribution of collision energies for the first collision or near miss in each integration. We show the distribution of collisions with a small enough distance of closest approach to result in a direct impact between the two cores (red), using the nominal core radii given by \citet{2013ApJ...770..131L}. We also show the distribution of near misses (green), defined as close encounters where $1<d_\mathrm{min}/(R_1+R_2)<1.2$.
The collision energy distribution exhibits long tails on both sides, with 55 collisions and near misses not shown in the scatterplot and collision-energy histogram, with minimum and maximum collision energies of $E_\mathrm{c}=-0.7E_\mathrm{b}$ and $E_\mathrm{c}=3.3E_\mathrm{b}$.
}
\label{Fig:Collision_Parameters}
\end{figure}

Figure~\ref{Fig:Collision_Parameters} shows the distribution of the first planet-planet collisions and near misses, defined as the first close encounter in each integration where the distance of closest approach is within $1.2$ times the sum of the planets' physical radii.
We characterize each collision and near miss by the distance of closest approach, $d_\mathrm{min}$, in units of the sum of the planets' physical radii, and the collision energy, $E_\mathrm{c}=E_\mathrm{k}+E_\mathrm{g}$ in units of the binding energy of both planets up to a coefficient, \begin{equation}E_\mathrm{b}=\frac{Gm_1^2}{R_1}+\frac{Gm_2^2}{R_2},\end{equation} where $E_\mathrm{k}$ and $E_\mathrm{g}$ are the kinetic and gravitational potential energy of the planets in the center of mass frame, ignoring the host star, where $R_1$ and $R_2$ are the radii of the planets.
We find that the number of initial collisions and near misses increases at higher distances of closest approach, and fit a linear probability density function, $p(d_\mathrm{min}) = 0.31d_\mathrm{min}+0.65$.
Adopting the densities reported in \citet{2013ApJ...770..131L}, we find that $37.3\%$ of the collisions result in a direct impact between the planet-cores and $20.7\%$ of the collisions result in a near miss where tidal effects could become important.
We find a significant difference ($p=7.7\times10^{-10}$) between the distributions of collision energy of first collisions compared to all collisions; specifically, subsequent collisions are in general more energetic than the first collision, likely due to the increased eccentricities required to generate subsequent orbit-crossings, consistent with \citet{2015ApJ...806L..26V}.
The collision energy qualitatively affects the outcome, specifically if the cores merge, the possibility of an ejection, and how much gas is retained by the remnant planet(s) \citep{2008ApJ...686..580C}.
Specifically, high values of both $d_\mathrm{min}$ and $E_\mathrm{c}$ lower the likelihood of a merger, and low values of both $d_\mathrm{min}$ and $E_\mathrm{c}$ increase the likelihood of a merger.

\begin{figure}[htp]
\begin{center}
\begin{tabular}{cc}
\includegraphics[width=8.0cm]{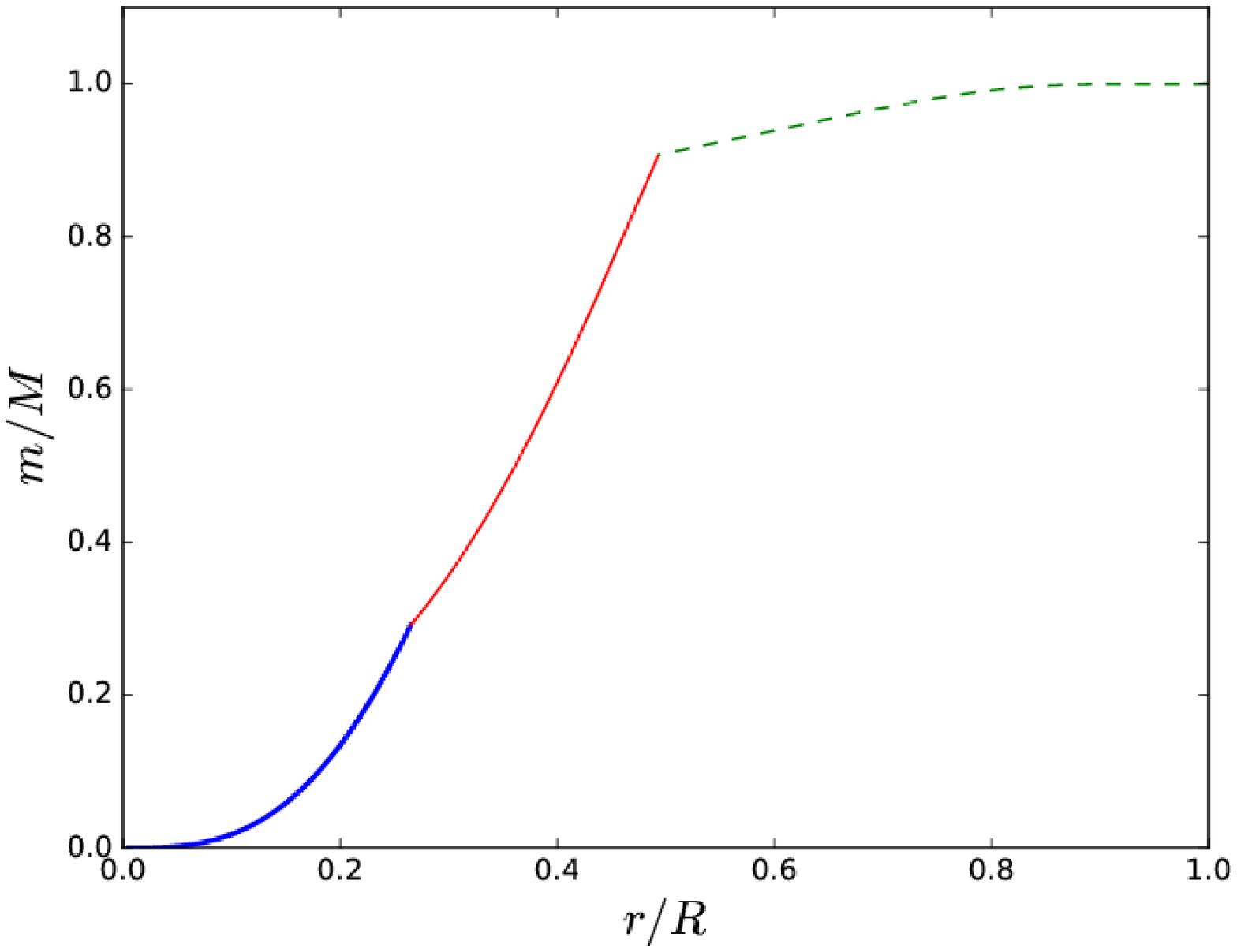}&
\includegraphics[width=8.0cm]{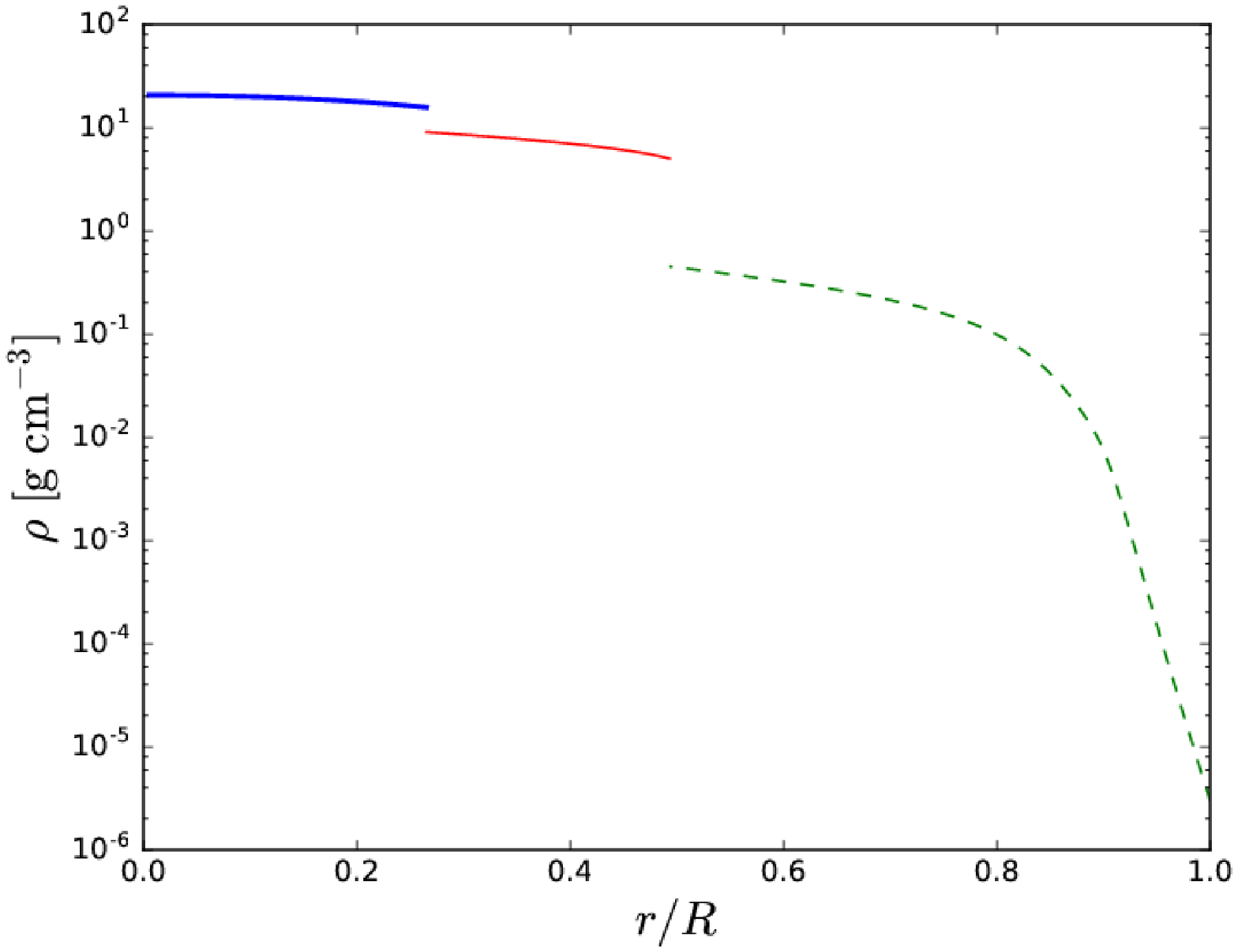}\\
\end{tabular}
\end{center}
\caption{Density (left) and enclosed mass (right) as a function of radius for a Kepler-11d analog with $6.6\%$ gas mass-fraction at $0.157$ AU.
We generate the gas envelope (dashed-green line) using MESA (see \S\ref{Sec:GasEnvelope}) and the core, made up of an inner iron core (thick-blue line) and a silicate mantle (red line), by integrating backwards from the envelope-core interface using the semi-analytic polytropic EoS from \citet{2007ApJ...669.1279S}.
We fit the iron core to silicate mantle mass-ratio by shooting for the desired core mass and radius.
The physical radius of our planet models depend most sensitively on the gas mass-fraction, consistent with \citet{2014ApJ...792....1L}.}
\label{Fig:Profiles}
\end{figure}

The details of planet-planet collisions are highly dependent on the radial profiles of the planets.
The structure of many planets found in Kepler multis is such that the mass, typically $3.0\ M_\oplus-9.0\ M_\oplus$, is dominated by a rocky (iron and/or silicate) core, but the volume is dominated by a tenuous gaseous atmosphere, which for collisions with a large enough distance of closest approach will likely result in gas being stripped from the planets while the cores remain intact.
Mass measurements either through radial velocity (RV) data or from transit timing variations (TTVs; \citealt{2005MNRAS.359..567A}; \citealt{2005Sci...307.1288H}; \citealt{2014ApJ...787...80H}) show that this generic two-component model is reasonable for the observed densities of the planets.
Figure~\ref{Fig:Profiles} shows the density and mass-profiles of a Kepler-11d analog, where we use the gas-mass fraction reported in \citet{2013ApJ...770..131L}, and we see that the gas envelope dominates the volume ($85\%$), but contributes only $6.6\%$ of the total mass.
We generate these mass-radius profiles using {\it Modules for Experiments in Stellar Astrophysics} (MESA; \citealt{2015ApJS..220...15P}) and equations of state from \citet{2007ApJ...669.1279S} (see \S\ref{Sec:GasEnvelope} and \S\ref{Sec:DiffCore} for more details).
The structure of sub-Neptunes, combined with the preference for grazing collisions, motivates us to look carefully at the validity and consequences of the often-used sticky-sphere approximation.
Hydrodynamic calculations are necessary to provide a better understanding of planet-planet collisions, such as the evolution of the stripped gas, specifically if the gas is reaccreted by the planets, falls into the star, or is ejected from the system.

\subsection{Hydrodynamic Calculations}
\label{SSec:SPH_Calculation}
Here we present hydrodynamic calculations of two representative collisions from our suite of dynamical integrations, using an SPH code, {\it StarSmasher}\footnote{{\it StarSmasher} is available at https://jalombar.github.io/starsmasher/.}, to treat the hydrodynamics.
We choose two integrations where Kepler-11d and Kepler-11e were the first planets to experience a collision, using the position and velocity coordinates of the planets many dynamical times prior to the close-encounter, where we treat the host star and the other planets in the system as point-mass particles that interact only gravitationally.
Table~\ref{TBL:Collision_Parameters} shows the characteristics of the two collisions; we purposefully choose two collisions where the degree of contact is high enough such that the physical cores do not come into contact, a barely-grazing collision, $d_\mathrm{min}=0.897(R_\mathrm{d}+R_\mathrm{e})$, and a deep collision, $d_\mathrm{min}=0.631(R_\mathrm{d}+R_\mathrm{e})$, where $R_\mathrm{d}$ and $R_\mathrm{e}$ are the radii of Kepler-11d and Kepler-11e.
We generate our gas-envelope profiles using {\it MESA} and use a differentiated, two-component model for our core, with an iron inner-core surrounded by a silicate mantle (see \S\ref{A:SPH_Models}), which may lead to slightly different overall densities, and thus planet radii, than those reported in \citet{2013ApJ...770..131L}.

\begin{deluxetable}{cccccccccc}
\tablewidth{16.5cm}
\tabletypesize{\footnotesize}
\tablecolumns{9}
\tablecaption{Collision and Planet Parameters \label{TBL:Collision_Parameters}}
\tablehead{
    \colhead{$Run$} & \colhead{$E_\mathrm{c}/E_\mathrm{b}$} & \colhead{$p(E_\mathrm{c}>E_{i})$} & \colhead{$d_\mathrm{min}/(R_d+R_e)$} & \colhead{$p(d_\mathrm{min}>d_{\mathrm{min},i})$} & \colhead{$R_d\ [R_\oplus]$} & \colhead{$R_e\ [R_\oplus]$} & \colhead{$(m_\mathrm{gas}/M)_d$} & \colhead{$(m_\mathrm{gas}/M)_e$} & \colhead{$\Delta E_\mathrm{tot}$}}
\startdata
$ 1 $ & $ 0.010 $ & $ 0.58 $ & $ 0.897 $ & $ 0.20 $ & $ 2.87 $ & $ 3.77 $ & $0.066$ & $0.157$ & $2.6\times10^{-4}$\\
$ 2 $ & $ 0.004 $ & $ 0.64 $ & $ 0.631 $ & $ 0.54 $ & $ 2.87 $ & $ 3.96 $ & $0.066$ & $0.157$ & $3.7\times10^{-5}$\\
\enddata

\tablenotetext{1}{Characteristics of the planet-planet collisions used in the SPH calculations, where $E_\mathrm{c}$ is the energy at collision in units of the binding energy of both planets, $p(E_\mathrm{c}>E_i)$ is the fraction of collisions with higher energies, $d_\mathrm{min}$ is the distance of closest approach, $p(d_\mathrm{min}>d_{\mathrm{min},i})$ is the fraction of collisions with higher distances of closest approach, $R_d$, $R_e$, $(m_\mathrm{gas}/M)_d$, and $(m_\mathrm{gas}/M)_e$ are the initial physical radii and gas mass-fractions of Kepler-11d and Kepler-11e, and $\Delta E_\mathrm{tot}$ is the fractional change in total energy.}
\end{deluxetable}

\begin{figure}[htp]
\begin{center}
\begin{tabular}{cc}
 \includegraphics[width=16.0cm]{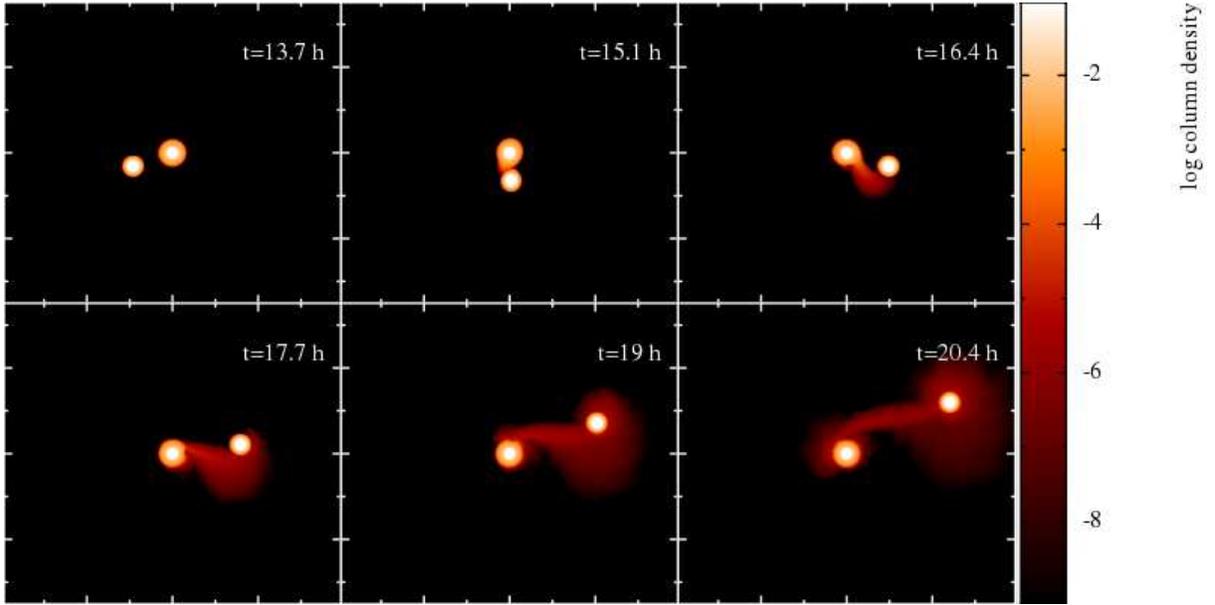}\\
\end{tabular}
\end{center}
\caption{
Logarithm of column density, in units of $M_\odot R_\odot^{-2}$, in the center of mass frame of Kepler-11e of an unstable system where Kepler11-d and Kepler-11e were the first planets to undergo a physical collision.
In this calculation $d_\mathrm{min}=0.897(R_\mathrm{d}+R_\mathrm{e})$, where due to the differences in the nominal radii and our gas-envelope profiles from {\it MESA}, results in a near miss, and the planets have nearly the same mass $q=0.95$.
We see a small amount of mass transferred from Kepler-11e to Kepler-11d due to tides, when the denser planet causes Roche-lobe overflow in the gas-rich planet.
We see that the collision clearly diverges from the sticky-sphere approximation, as both planets survive the close encounter.}
\label{Fig:Run1}
\end{figure}

\begin{figure}[htp]
\begin{center}
\begin{tabular}{cc}
 \includegraphics[width=16.0cm]{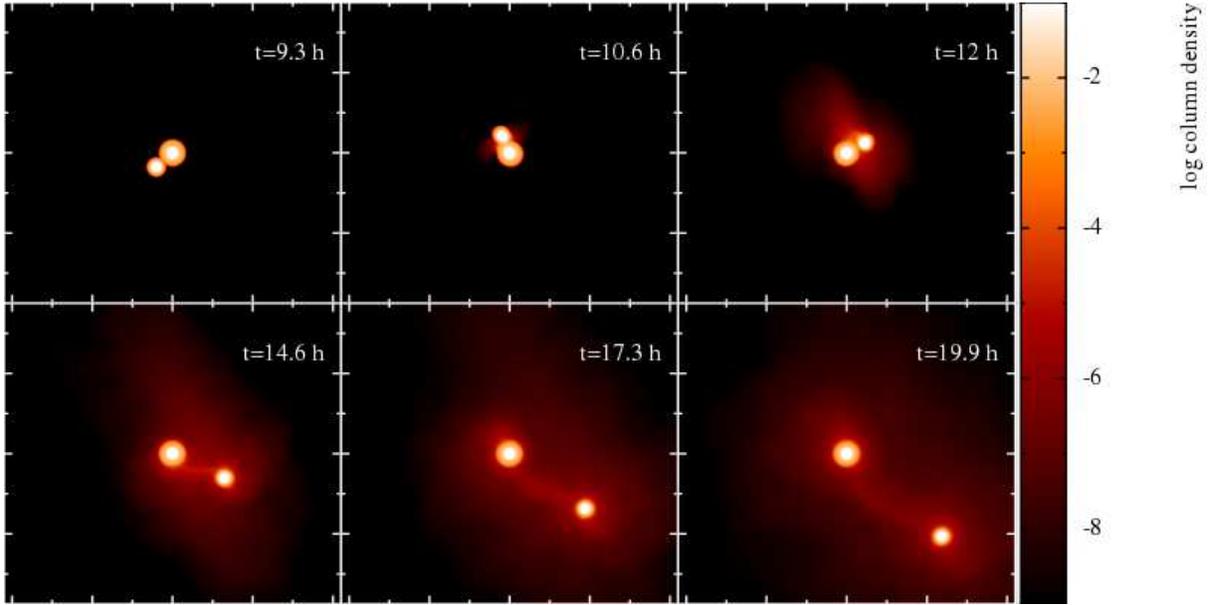}\\
\end{tabular}
\end{center}
\caption{
Logarithm of column density, in units of $M_\odot R_\odot^{-2}$, in the center of mass frame of Kepler-11e of an unstable system where Kepler11-d and Kepler-11e were the first planets to undergo a physical collision.
In this calculation $d_\mathrm{min}=0.631(R_\mathrm{d}+R_\mathrm{e})$, a deep collision, with a mass ratio, $q=0.53$.
We see that both envelopes are significantly disrupted during the collision, with mass-transfer from Kepler-11d to Kepler-11e.
Most of the disrupted-gas remains tenuously bound to the planets, with less than $0.1\%$ of the gas-mass ejected from both planets.
We see that the collision clearly diverges from the sticky-sphere approximation, as both planets survive the close encounter.}
\label{Fig:Run2}
\end{figure}

Figures~\ref{Fig:Run1} and \ref{Fig:Run2} show the time evolution of the hydrodynamic calculations (see \S\ref{A:SPH_Models} for details), and we see that in both cases the planets survive the initial contact.
Table~\ref{TBL:Collision_Results} shows the change in mass and orbits of Kepler-11d and Kepler-11e after the collision.
In Run 1, we see the effect of different gas-mass fractions on the structure of two nearly equal mass planets ($q=0.95$), where initially the gas-rich planet, Kepler-11e, has over twice the volume of the gas-poor planet, Kepler-11d.
In this calculation there is minimal contact between the two planets, resulting in a small amount of mass transfer due to Kepler-11d tidally disrupting the outer envelope of Kepler-11e.
We also find that, after the collision the orbits, orbital energy, and orbital angular momentums are flipped, where Kepler-11d has a final semi-major axis larger than Kepler-11e.
Run 2 shows a much deeper, more energetic collision between two planets with a mass ratio $q=0.53$, where we see significant disruption in both envelopes and a small amount of mass transfer from Kepler-11e to Kepler-11d.
In both calculations we see a clear deviation from the sticky-sphere approximation, and although we cannot rule out the result that the two planets will eventually collide again, due to the exchange and loss of angular momentum and energy there exist collisions that result in two surviving planets in a more stable configuration.
These results motivate developing a more accurate prescription for handling collisions in N-body integrations, including mass-transfer between planets and ejection and loss of mass in merger remnants.

\begin{deluxetable}{ccccccccc}
\tablewidth{12.0cm}
\tabletypesize{\footnotesize}
\tablecolumns{8}
\tablecaption{Collision Results \label{TBL:Collision_Results}}
\tablehead{
    \colhead{$Run$} & \colhead{$Planet$} & \colhead{$m_i\ [M_\oplus]$} & \colhead{$\Delta m\ [M_\oplus]$} & \colhead{$a_i\ [AU]$} & \colhead{$a_f\ [AU]$} & \colhead{$e_i\ [AU]$} & \colhead{$e_f\ [AU]$}}
\startdata
$ 1 $ & $ Kepler-11d $ & $ 7.23 $ & $ 0.005 $ & $ 0.156 $ & $ 0.177 $ & $ 0.173 $ & $ 0.110 $\\
$ 1 $ & $ Kepler-11e $ & $ 7.58 $ & $-0.005 $ & $ 0.178 $ & $ 0.157 $ & $ 0.140 $ & $ 0.182 $\\
$ 2 $ & $ Kepler-11d $ & $ 5.39 $ & $ 0.003 $ & $ 0.151 $ & $ 0.149 $ & $ 0.076 $ & $ 0.056 $\\
$ 2 $ & $ Kepler-11e $ & $10.25 $ & $-0.011 $ & $ 0.156 $ & $ 0.157 $ & $ 0.046 $ & $ 0.055 $\\
\enddata

\tablenotetext{1}{Initial and change in the planets' masses, $m$, and initial and final semi-major axes, $a$, and eccentricities, $e$, of our hydrodynamic calculations after 1000 dynamical times.}
\end{deluxetable}

\section{Consequences of Planet-Planet Collisions}
\label{Sec:Consequences}

\subsection{Observational Signatures of Mergers}
\label{SSec:MergerRemnants}
\begin{figure}[htp]
\begin{center}
\begin{tabular}{cc}
\includegraphics[width=8.0cm]{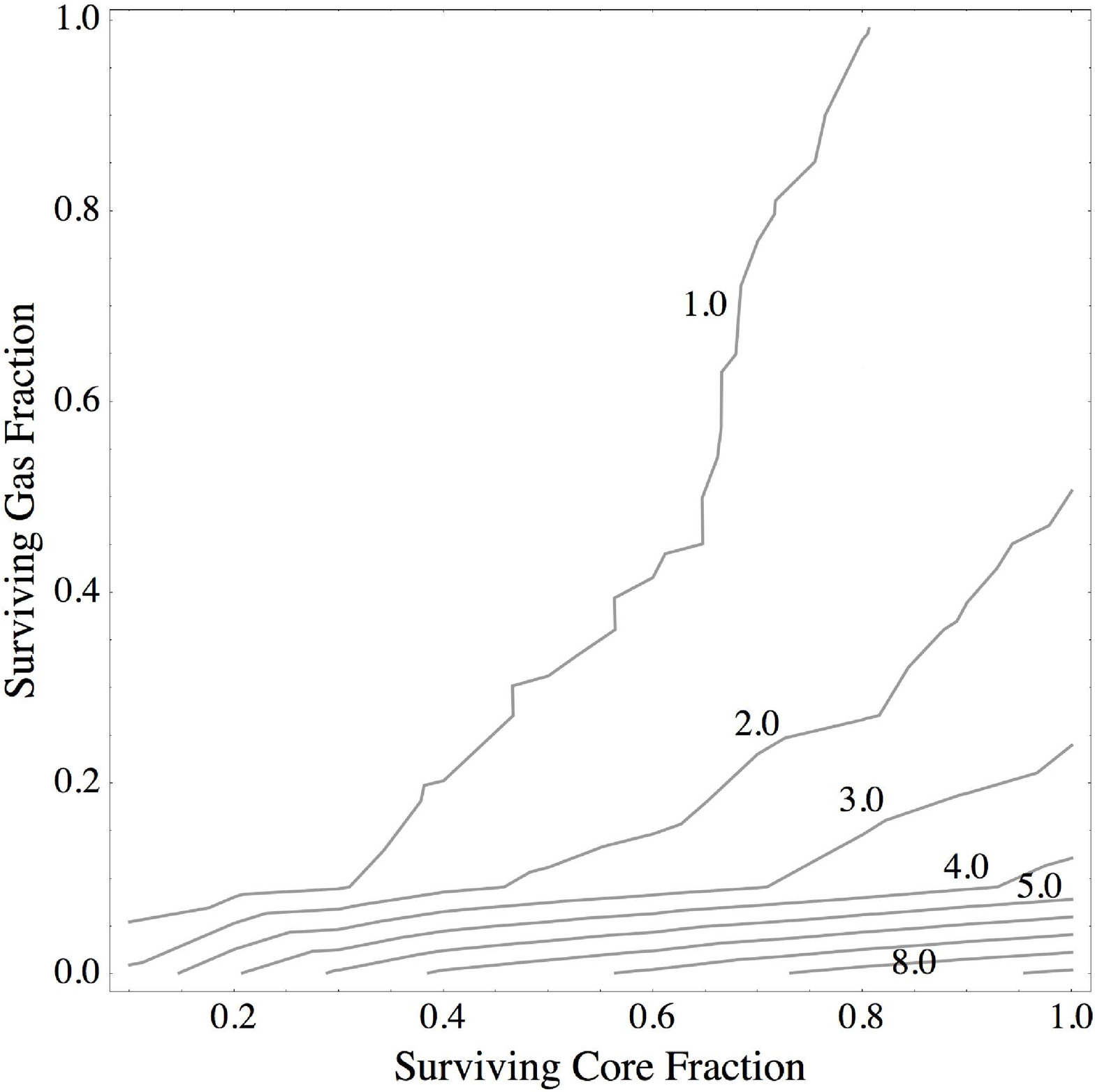}
\end{tabular}
\end{center}
\caption{Density of the merger remnant resulting from the collision of two identical 7 $M_\oplus$ planets with 15\% of their mass in a gaseous envelope as a function of the fraction of retained core and gas material, where we include the mantle with the core.  For planets where a large fraction of core material is retained and a small amount of gas, the planet produced by the collision has roughly double the density of the ingoing planets (with predicted densities of 4.1 g/cm$^3$).}
\label{equalmass}
\end{figure}

Assuming that a collision results in a merger, a sizable fraction of the gas and, perhaps, of the mantle, can be stripped away \citep{2015MNRAS.448.1751I}.
This preferential removal of the lighter material leaves the remnant with a larger mass, but a similar or somewhat smaller radius.
The amount of light material stripped depends sensitively on the collision velocity, distance of closest approach, and mass-profile of the planets.
Using a prescription from \citet{2014ApJ...792....1L}, we consider the steady-state outcomes of these collisions from a phenomenological standpoint.
Figure \ref{equalmass} shows the resulting densities of collisions between two identical $7.0\ M_\oplus$ planets with 15\% of their mass in a H/He envelope as a function of the core and gas mass-fractions of the remnant.
We see that as the fraction of residual gas declines following the collision, the density of the resulting planet grows sharply.
Therefore, should collisions generally remove the bulk of the gas from the planets, the result would be a planet with an anomalously large density.

\begin{figure}[htp]
\begin{center}
\begin{tabular}{cc}
\includegraphics[width=8.0cm]{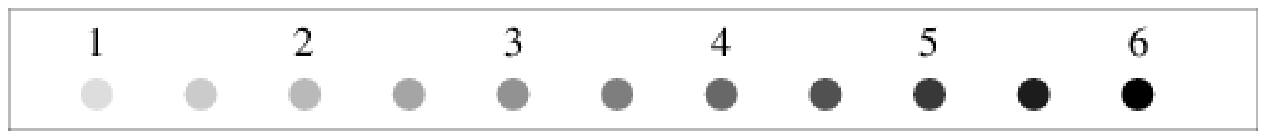}\smallskip \\
\includegraphics[width=8.0cm]{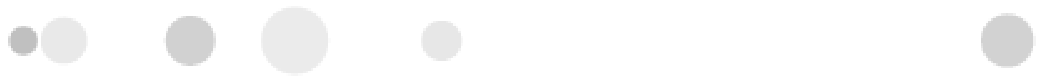}\smallskip \\
\includegraphics[width=8.0cm]{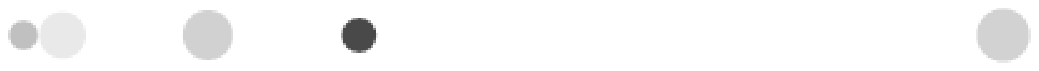}\smallskip \\
\includegraphics[width=8.0cm]{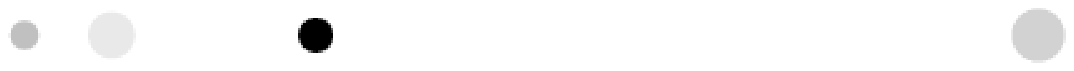}\smallskip \\
\includegraphics[width=8.0cm]{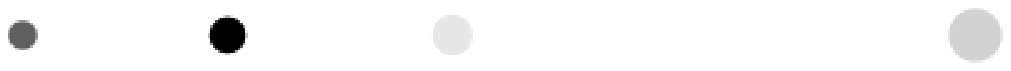}\smallskip \\
\includegraphics[width=8.0cm]{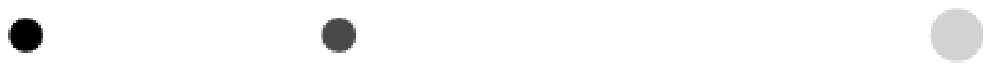}\smallskip \\
\includegraphics[width=8.0cm]{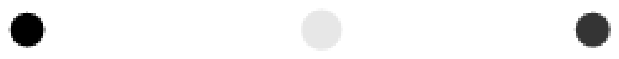}\smallskip \\
\end{tabular}
\end{center}
\caption{Outcomes of unstable Kepler-11 realizations using the parameters taken from \citet{2013ApJ...770..131L}.
The disks represent planets located at their relative semi-major axes.
The size of the disk is proportional to the radius of the planet, where the disks in the top row are scaled to two Earth radii.
The shading of the disks give the planet density in grams per cubic centimeter.
The second row is the canonical Kepler-11 system while the remaining five rows are the remnants of chosen unstable dynamical integrations.
For each collision we assume that the remnant retains $95\%$ of the core material and $5\%$ of the gaseous envelope, resulting in gas-depleted merger remnants that have a higher density than the planets that did not undergo a merger.}
\label{Fig:Kepler_11_Densities}
\end{figure}

We use the best fitting two-component planet models from \citet{2013ApJ...770..131L} to look at the masses and densities of the planets in the system after one or more collisions.
As an example, we fix the remnant core mass to be $95\%$ of the initial core masses and the remnant envelope mass to be $5\%$ of the initial envelope masses.
Figure \ref{Fig:Kepler_11_Densities} shows several examples of the outcomes, displaying both the sizes and densities of the remaining planets, where the mean density of the ingoing planets is roughly $4.1$ g/cm$^3$.
We see that the typical size of the resulting remnant planet is similar to those of the two injected planets, in some cases even slightly smaller due to the large amount of gas removed from the system.
The density variations are much more prominent---colliding planets with initial densities of $1-2$ g/cm$^3$ produce remnants with densities of $5-6$ g/cm$^3$.
This large change in density is an observable consequence of post-disk planet-planet collisions where no remaining gas from the protoplanetary disk is available to replenish the lost gaseous material.

Aside from the differences in densities, another effect of collisions is that in general the remnant planet is more isolated from its neighbors, with an overall increase in the dynamical separation.
For systems consisting of closely spaced sub-Neptunes formed in situ or driven to their locations through roughly homologous migration, one would generally expect the most isolated planet to be among the largest since it would have had the most available material to accrete.
Therefore, an isolated, high-density planet could provide a signature of a previous planet-planet collision.
Another byproduct of post-disk collisions is a bimodal mass-radius relationship, the primary mode being the generic sub-Neptune planets and the secondary mode being the gas depleted counterparts with much higher densities.
The relative numbers of planets in the two populations may provide an estimate for the fraction of systems that underwent previous planet-planet collisions.

Two effects yielding potential differences in the sizes of the remnant planets include variations in incident flux and inflation of the gas envelope of the collision remnant(s).
The incident flux varies significantly between systems and will be important when modeling the gas envelope in detailed hydrodynamic calculations; we use {\it MESA} to generate our irradiated planet profiles.
The gas envelope of the remnant planet(s) will be inflated after the collision; the typical cooling time for the planets used in our analysis is of order 10 million years and we predict that similar cooling times will apply to the post-collision planets.
We expect observations of inflated atmospheres resulting from collisions to be relatively rare since the cooling time is much less than the age of the system, and a collision remnant with an inflated atmosphere would not likely draw attention since it would appear larger than its steady-state size, rendering the observed density more consistent with the other planets in the same system.

\begin{figure}[htp]
\begin{center}
\begin{tabular}{cc}
 \includegraphics[width=8.0cm]{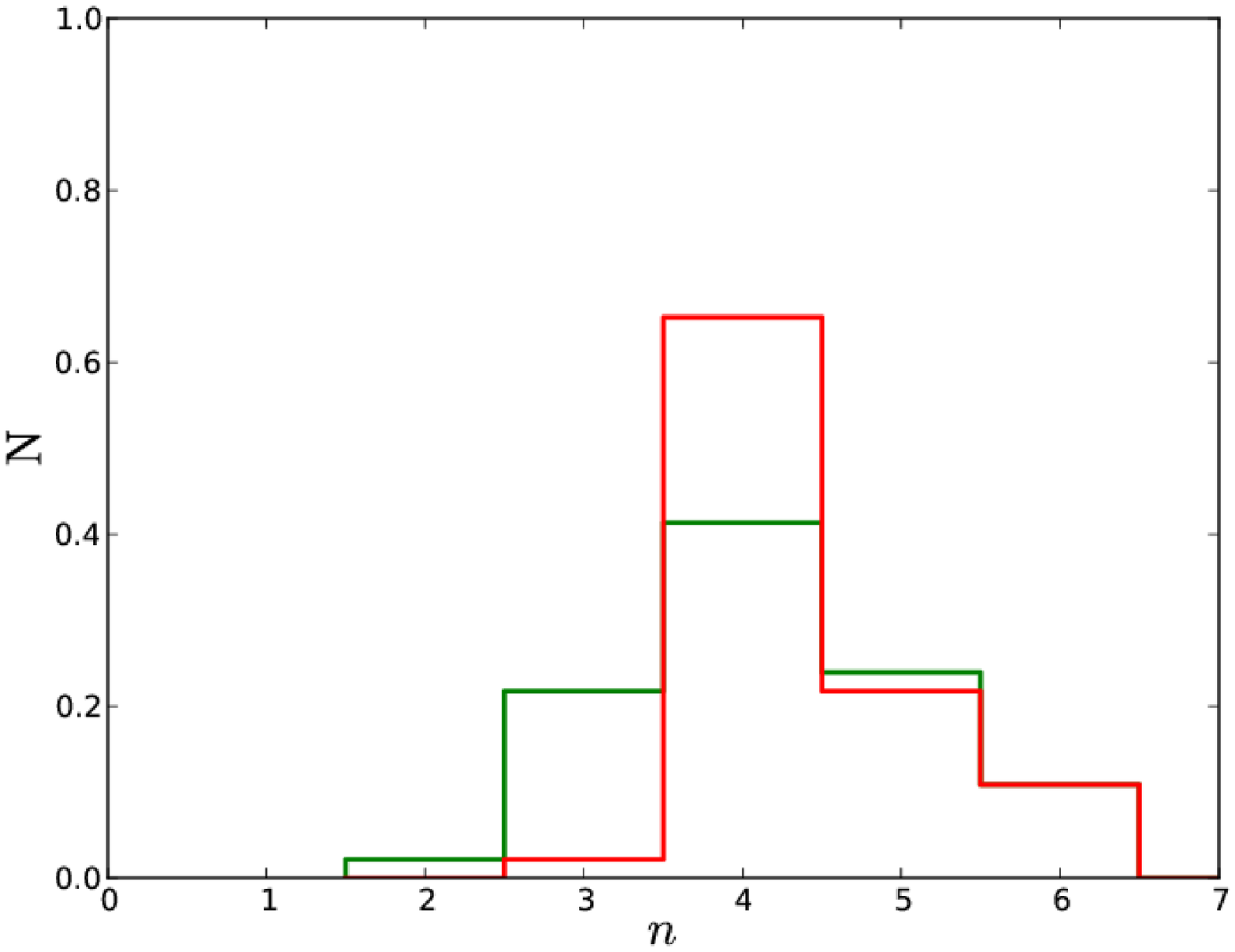}&
 \includegraphics[width=8.0cm]{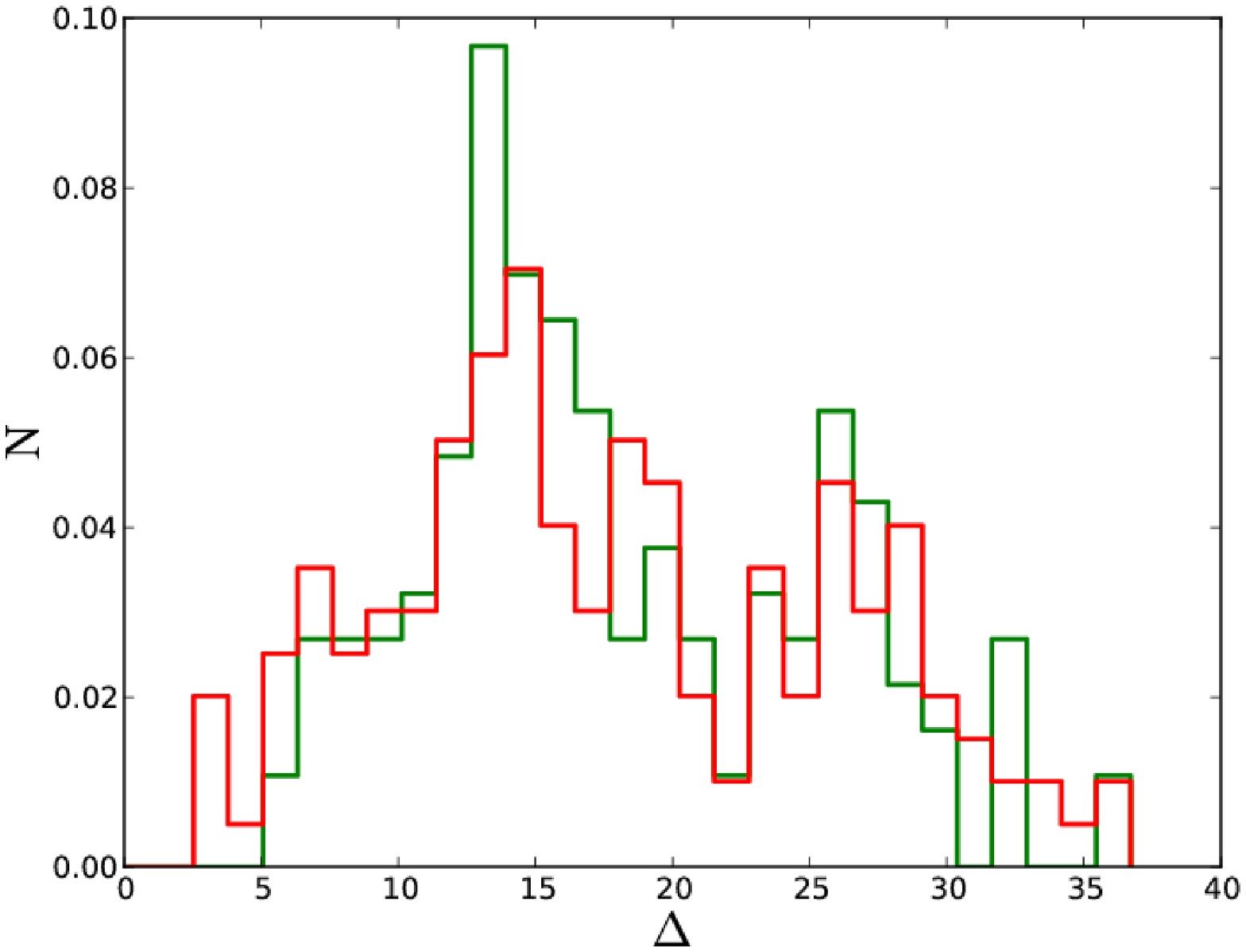}\\
\end{tabular}
\end{center}
\caption{Multiplicity (left) and dynamical separation (right) distributions of the remnant systems from two sets of dynamical integrations using the same initial conditions, where one uses the sticky-sphere approximation (green), and the other uses our modified collision prescription (red).
We find that, in general, the integrations using our modified collision prescription have higher remnant multiplicities.}
\label{Fig:Sticky_Sphere}
\end{figure}

\subsection{Modified N-Body Calculations}
\label{Sec:NBody2}
Here, we show the results of a simplistic modified prescription for collisions in a subset of the dynamical integrations presented in \S\ref{Sec:Results}.
In our simple prescription, close encounters with a distance of closest approach greater than the assigned core radii immediately eject their envelopes without merging, and we update the density and mass accordingly.
Close encounters with a distance of closest approach less than the size of their physical cores are treated with the existing sticky-sphere approximation.
In the limiting case where the physical core is a point mass the system will never experience any mergers; instead, close encounters with a small enough distance of closest approach will result in either planets infalling into the host star or ejections.
Figure~\ref{Fig:Sticky_Sphere} compares the multiplicity and dynamical separation distributions between two sets of integrations with identical initial conditions, one with the original sticky-sphere prescription and the other with our modified prescription.
We see a significant increase in the number of four-planet remnant systems, likely because the mass loss from the ejected gas increases the dynamical spacing between planets after a close encounter, reducing the likelihood of subsequent close encounters. 
We also observe a broader spread in the dynamical separation, likely due to the higher multiplicity of the remnants.

\section{Conclusions and Discussions}
\label{Sec:Conclusions}

\subsection{Summary of Results and Discussion}
We conduct a detailed study on the dynamical evolution of high-multiplicity, tightly-packed systems, using Kepler-11 as a nominal example to generate our systems.
We find that the fraction of systems that remain stable as a function of time is well described by a power-law.
We use a $\chi^2$-test on the hypothesis that the stability of the system is independent of the masses and orbital parameters and planet-pair characteristics, and summarize our findings as follows:
\begin{itemize}\setlength{\itemsep}{-4pt}
  \item[1)]Of the masses and orbital elements, the stability of the system is least likely to be independent from the eccentricities of the planets, and we find statistically significant peaks in stability at $e>0$ due to a lower oscillation amplitude.
  \item[2)]The stability of the system is unlikely to be independent from every planet-pair characteristic we examined (magnitude of action-angle variables, dynamical separation, distance to MMR, and period ratio), with much lower p-values than any individual mass or orbit parameter.
  \item[3)]The stability of the system greatly increases with dynamical separation up to a point, and no system was stable with a $\Delta<4.93$.
  \item[4)]The stability of the system is least likely to be independent from the magnitude of the action-angle variables, $R$, defined in \eqref{EQ:R}, and the percentage of stable systems decreases sharply as $R$ increases. Excluding Kepler-11fg, no stable system had a planet pair with $R>6.2$.
  \item[5)]We find a statistically significant trough at $\epsilon=0$ and statistically significant peaks both inside and outside of $\epsilon=0$, particularly at the 3:2 period ratio, suggesting that chaos near resonances may drive instability.
\end{itemize}

We examine in detail the architecture of unstable systems in general, as well as in subsets organized by remnant multiplicity.
We find an overall increase in eccentricity and mutual inclination for systems with an instability, and that after a merger the dynamical separations become more evenly-spaced.
We generate a synthetic population of observed systems from our dynamical integrations and find reasonable agreement with the observed Kepler sample when comparing distributions of dynamical separation, supporting our hypothesis that many of the observed Kepler multis may be remnants of a population of even higher-multiplicity systems that have undergone post-disk instabilities.

We find that most collisions tend to have a relatively large distance of closest approach, which, depending on the composition of the planets, suggests a low rate of direct contact, at least initially, between the physical cores.
We present two representative calculations between Kepler-11d and Kepler-11e from our suite of dynamical integrations, one barely-grazing and one deep collision.
In both cases we observed mass transfer, mass ejected from the system, and both planet surviving the initial collision.
While this does not guarantee that the planets will avoid future close encounters, the change in both orbital energy and angular momentum dictate that there do exist collisions that stabilize the two planets.

We use a prescription from \citet{2014ApJ...792....1L} to predict observable properties of the merger remnants and propose a bimodal mass-radius relationship in the observed Kepler systems, with the primary mode being the generic sub-Neptune planets and the secondary mode being the gas-depleted, high-density remnants resulting from planet-planet collisions.
We rerun a subset of our dynamical integrations using a modified prescription for collisions and see divergence in the orbital evolution of many realizations; specifically, we see collisions occurring between different planets and in general, higher-multiplicity remnant systems.

The results we have presented here agree qualitatively with those of other recent studies.
\citet{2015ApJ...807...44P} conduct a detailed study of stability in high-multiplicity, non-circular, and non-coplanar systems and report the effects of eccentricity and inclination on the dynamical separations required to maintain stability for a given timescale.
In our work, because we draw our models from a nominal system that experiences multiple types of instability (Fig.~\ref{Fig:t_first_instability} exhibits the scatter about $\Delta_\mathrm{min}$, a commonly used metric to measure stability), we focus on the sensitivity of stability on the masses, orbital parameters, and various planet-pair characteristics.
\citet{2015ApJ...806L..26V} integrate many analogs of tightly-packed systems, including Kepler-11, and characterize the first collisions in each integration compared to subsequent collisions, and we agree with their result that subsequent collisions are more energetic than the first collisions.

This research was supported in part through the computational resources and staff contributions provided for the Quest high performance computing facility at Northwestern University, which is jointly supported by the Office of the Provost, the Office for Research, and Northwestern University Information Technology.
FAR and JAH were supported by NASA Grant NNX12AI86G to FAR.
JHS was supported by a grant from the Kepler Participating Scientist Program (NNH12ZDA001N-KPS) and through the Lindheimer Fellowship at Northwestern University.
JCL was supported by NSF Grant AST-1313091.
We thank Joshua Fixelle for useful discussions while developing the equations of state used in the SPH calculations and Francesca Valsecchi for help with using {\it MESA} to generate sub-Neptune envelopes.
We thank the referee for providing helpful feedback leading to improvements in this manuscript.
This work used the SPLASH visualization software \citep{2007PASA...24..159P}.

\pagebreak
\appendix

\section{Appendix}

\subsection{Determining Transits}
\label{A:Transit}
To generate a synthetic 'observed' population from our dynamical integrations we use the final orbital elements from our dynamical integrations to determine which planets transit with respect to a randomly generated line of sight.
We sample each dynamical integration 300 times with a random line of sight vector, \begin{equation}\mathbf{p}(s)=s\begin{pmatrix}\sqrt{1-u^2}\cos(\theta)\\\sqrt{1-u^2}\sin(\theta)\\u\end{pmatrix},\end{equation} where $s\sim r$ and we randomly sample $u\in[0,1)$ and $\theta\in[0,2\pi)$.
We parameterize the orbit as \begin{equation}\mathbf{q}(t)=r(t)\begin{pmatrix}\cos(\Omega)\cos(\omega+f)-\sin(\Omega)\sin(\omega+f)\cos(i)\\\sin(\Omega)\cos(\omega+f)+\cos(\Omega)\sin(\omega+f)\cos(i)\\\sin(\omega+f)\sin(i)\end{pmatrix},\end{equation} where \begin{equation}r(t)=a(1-e\cos(t))\end{equation}
is the orbital radius, \begin{equation}f(t)=2\tan^{-1}\left(\sqrt\frac{1+e}{1-e}\tan\left(\frac{t}{2}\right)\right)\end{equation}
is the true anomaly, $a$ is the semi-major axis, $e$ is the eccentricity, $i$ is the inclination, $\omega$ is the argument of pericenter,  $\Omega$ is the longitude of ascending node, and $t$ is the eccentric anomaly \citep{1999ssd..book.....M}.
We find the minimum distance, $d=\mathrm{min}(|\mathbf{p}(s)-\mathbf{q}(t)|),$ and determine that the planet transits if \begin{equation}r\sin(d/r)\le R_*.\end{equation}

\subsection{Predicting Stability}
\label{A:Bayes}
We use Bayes' Theorem to calculate the likelihood that a system will remain stable until some time $t_\mathrm{f}$ given that it is stable at $t_\mathrm{o}$,
\begin{equation}P(t_\mathrm{f}|t_\mathrm{o})=\frac{P(t_\mathrm{o}|t_\mathrm{f})P(t_\mathrm{f})}{P(t_\mathrm{o})}
\end{equation}
We fit the fraction of systems that are stable as a power law, $f(t)$, in \eqref{EQ:PowerLaw} and estimate $P(t)\simeq f(t)$.
As $P(t_\mathrm{o}|t_\mathrm{f})=1$, we have \begin{equation}P(t_\mathrm{f}|t_\mathrm{o})\simeq
\begin{cases}
      f(t_\mathrm{f})/f(t_\mathrm{o}) & t_\mathrm{o}<t_\mathrm{f}\\
      1 & t_\mathrm{o}\ge t_\mathrm{f}\\
\end{cases}.\end{equation}
We define the lowest integration time of a stable system as $t_\mathrm{c}$, and for each system that has a first instability at $t>t_\mathrm{c}$, we project the fraction of systems that are stable at this time as \begin{equation}n_\mathrm{S}(t)=\frac{1}{N(t)+M(t)}\sum\limits_{i=1}^{N}P(t_\mathrm{f}=t|t_\mathrm{o}=t_i),\end{equation} where $N$ is the number of systems that are stable, $M$ is the number of systems that are unstable, and $i$ iterates over all the systems that are stable (see Figure~\ref{Fig:t_integrations} at time $t$).

\subsection{Creating SPH Models of sub-Neptunes}
\label{A:SPH_Models}
In this section we describe how we generate realistic planet models for the SPH calculations.
We generate the gas-envelope profiles using {\it MESA} and fit a differentiated core, made up of an inner iron-core surrounded by a silicate mantle, based on the inner boundary conditions given by {\it MESA}.
We fit an equation of state for the gas envelope from a grid of models generated using {\it MESA} and supply a semi-analytic polytropic equation of state, fit to empirical data, for the iron core and silicate mantle.
Finally, we describe how we handle the abrupt transitions between the gas envelope, silicate mantle, and inner iron-core.

\subsubsection{Generating a Gas Envelope}
\label{Sec:GasEnvelope}
We use {\it MESA} (\citealt{2011ApJS..192....3P}; \citealt{2013ApJS..208....4P}; \citealt{2015ApJS..220...15P}) to generate gaseous envelopes with a constant-density core of mass $m_\mathrm{c}$ using the following steps (adapted from \citealt{2013ApJ...769L...9B}):
\begin{itemize}\setlength{\itemsep}{-4pt}
  \item[1)]Relax a gaseous cloud of mass $M_\mathrm{i}=0.2M_\mathrm{J},$ a mass {\it MESA} is able to initially generate at equilibrium.
  \item[2)]Remove mass equal to the core mass, $m_\mathrm{c},$ and place a hard-sphere boundary at the radius where a constant density core of mass $m_\mathrm{c}$ would be, where the density is chosen, assuming a differentiated core of $67\%$ Silicate Mantle and $33\%$ Iron by mass, from interpolating a table provided in \citet{2014ApJ...787..173H}.
  \item[3)]Remove mass from this model slowly enough so that the model remains in equilibrium until the desired planet mass is reached.
  \item[4)]Irradiate the planet at the chosen semi-major axis, $a$, and column depth for irradiation, $\Sigma_*=2/\kappa_v$, where we choose an opacity,  $\kappa_v=10^{-2}\ \mathrm{cm}^2\ \mathrm{g}^{-1}$.
\end{itemize}

\subsubsection{Generating a Core}
\label{Sec:DiffCore}
We replace the isothermal and constant-density core {\it MESA} uses as a boundary condition with a core generated from an appropriate equation of state.
For our models we generate a differentiated core consisting of an iron core surrounded by a silicate mantle.
We choose to use the semi-analytic polytropic equations of state provided by \citet{2007ApJ...669.1279S}, \begin{equation}\label{EQ:CoreEoS}p = c(\rho-\rho^\prime)^\gamma,\end{equation} where $\rho^\prime$, $c$, and $\gamma$ are defined by the composition (e.g. $MgSiO_3$ or $Fe$) and are fit to more detailed equations of state, specifically Thomas-Fermi-Dirac (TFD) at $P>10^4\ \mathrm{GPa}$ \citep{1967PhRv..158..876S}, empirical fits to data via the Vinet (\citealt{1987PhRvB..35.1945V}; \citealt{1989JPCM....1.1941V}) and Birch-Murnagham \citep{1947PhRv...71..809B} equations of state at $P<200\ \mathrm{GPa}$. The cutoff point to switch from the low-pressure regime to the high-pressure regime occurs at the intersection between the two equations of state.

We numerically integrate backwards using the pressure at the inner boundary condition provided by {\it MESA} and the core mass, enforcing hydrostatic equilibrium, \begin{equation}\frac{dp}{dr}=-\frac{GM(r)\rho(r)}{r^2},\end{equation} where $M(r)$ and $\rho(r)$ are the enclosed-mass and density at a radius $r$.
We first check whether the mass and radius of the core are consistent with our assumptions by integrating both a completely iron core and a completely silicate mantle to check if the core radii are too small and too large, respectively.
We begin with an initial guess that the iron core is $33\%$ of the total core-mass and shoot for the mass-ratio that returns the mass and radius of the core from the {\it MESA} model.
We generate an internal energy profile assuming an adiabatic process, \begin{equation}du=\frac{p}{\rho^2}d\rho,\end{equation} yielding \begin{equation}u=\frac{c\rho^{\gamma-1}{}_2F_1\left(1-\gamma,-\gamma,2-\gamma,\frac{\rho^\prime}{\rho}\right)}{\gamma-1}\label{EQ:Seager_u},\end{equation} where ${}_2F_1(1-\gamma,-\gamma,2-\gamma,\frac{\rho^\prime}{\rho})$ is the ordinary hypergeometric function.

\subsubsection{Equations of State}
\label{Sec:EoS}
\begin{figure}[htp]
\begin{center}
\begin{tabular}{cc}
\includegraphics[width=17.0cm]{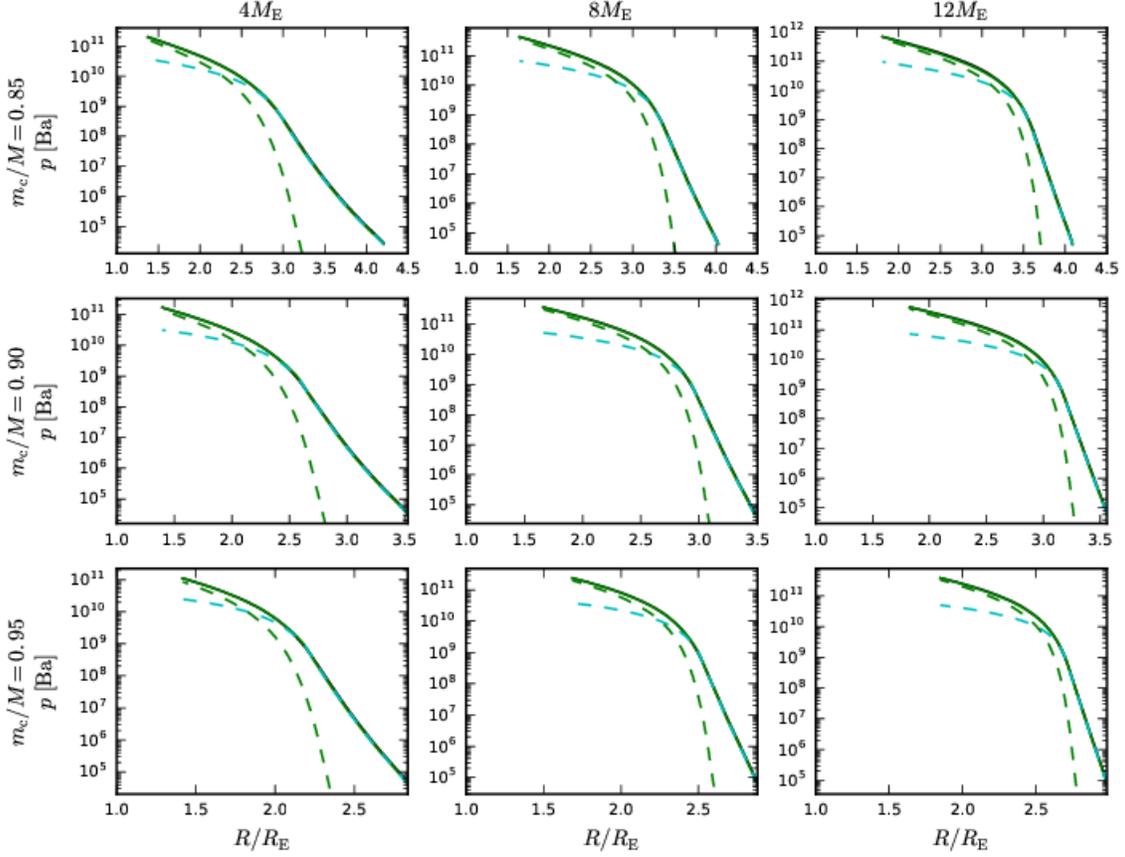}\\
\end{tabular}
\end{center}
\caption{Pressure profiles for a grid of gas envelopes generated by {\it MESA}, showing $4.0M_\oplus$ (left column), $8.0M_\oplus$ (middle column), and $12.0M_\oplus$ (right column) planets with core mass fractions $m_\mathrm{c}/M=0.85$ (first row), $m_\mathrm{c}/M=0.90$ (second row), and $m_\mathrm{c}/M=0.95$ (third row).
The {\it MESA} profile (solid black) agrees very well with our fit (solid green) consisting of a polytrope (dashed green) with $\gamma_\mathrm{e}=3$ and ideal gas (dashed cyan), detailed in \eqref{EQ:pGasEoS}.}
\label{Fig:FitEoSp}
\end{figure}

\begin{figure}[htp]
\begin{center}
\begin{tabular}{cc}
\includegraphics[width=17.0cm]{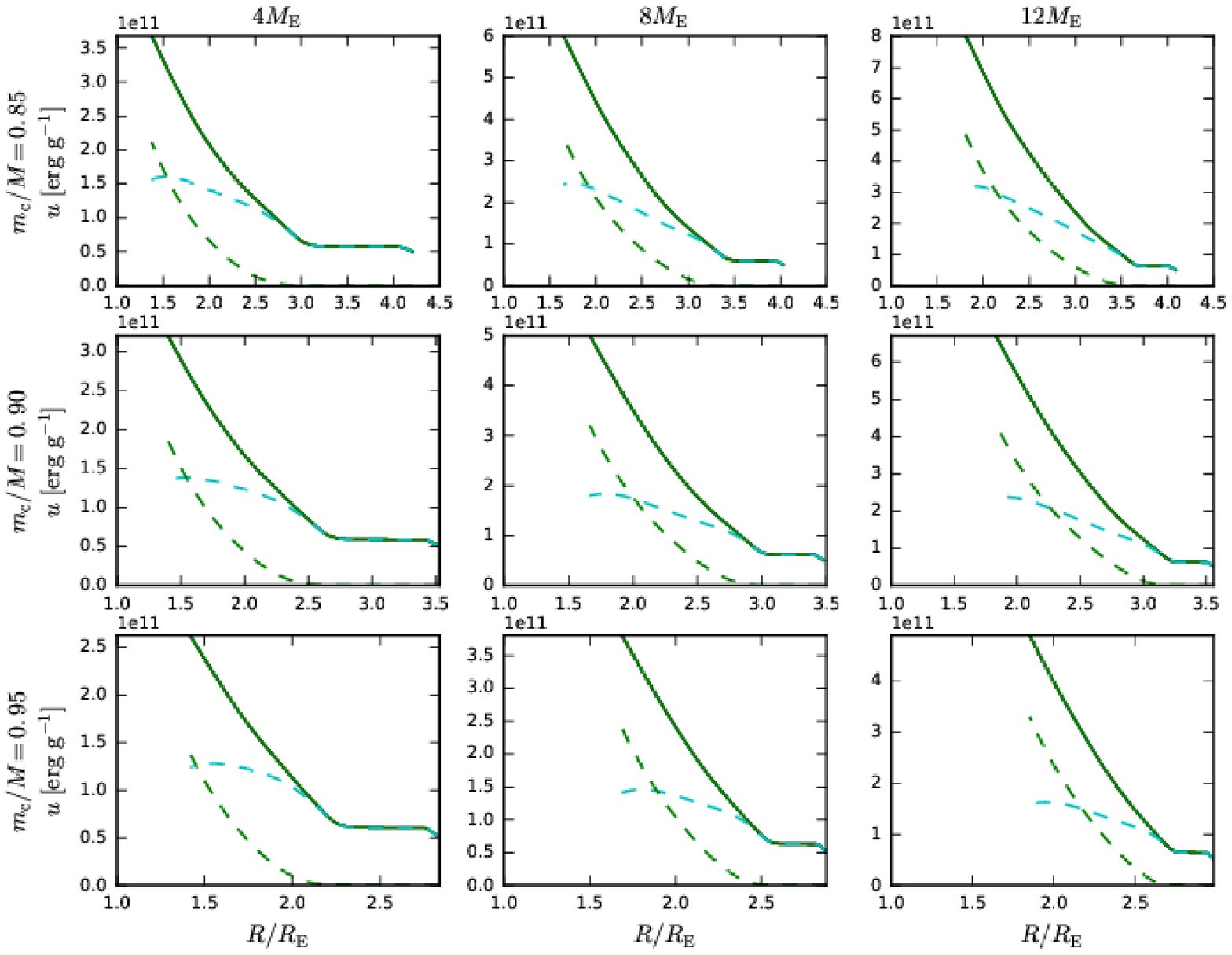}\\
\end{tabular}
\end{center}
\caption{Internal energy profiles for a grid of gas envelopes generated by {\it MESA}, showing $4.0M_\oplus$ (left column), $8.0M_\oplus$ (middle column), and $12.0M_\oplus$ (right column) planets with core mass fractions $m_\mathrm{c}/M=0.85$ (first row), $m_\mathrm{c}/M=0.90$ (second row), and $m_\mathrm{c}/M=0.95$ (third row).
The {\it MESA} profile (solid black) agrees very well with our fit (solid green) consisting of a polytrope (dashed green) with $\gamma_\mathrm{e}=3$ and ideal gas (dashed cyan), detailed in \eqref{EQ:uGasEoS}.}
\label{Fig:FitEoSu}
\end{figure}

We supply {\it StarSmasher} with an equation of state to calculate the pressure dependent on the density, internal energy, composition, and mean molecular weight of each particle.
Particles in the core are treated with the semi-analytic polytropic equations of state \eqref{EQ:CoreEoS} and are assigned a composition dependent on their initial position.
For the gas particles we calculate the pressure and internal energy using an equation of state, \begin{equation}\label{EQ:pGasEoS}p_i=K_\mathrm{e}\rho_i^{\gamma_\mathrm{e}}+\frac{\rho_ik_\mathrm{B}T_i}{\mu_im_H}\end{equation} and
\begin{equation}\label{EQ:uGasEoS}u_i=\frac{K_\mathrm{e}\rho_i^{\gamma_\mathrm{e}-1}}{\gamma_\mathrm{e}-1}+\frac{\beta_ik_\mathrm{B}T_i}{\mu_im_H},\end{equation} where the polytropic term approximates the electrostatics, $\rho_i$, $T_i$, and $\mu_i$ are the density, temperature, and mean molecular weight of the particle $i$, $\beta_i$ corrects for the degrees of freedom, $k_\mathrm{B}$ is the Boltzmann constant, and $m_\mathrm{H}$ is the mass of a Hydrogen atom.
We empirically determine the value of $\gamma_\mathrm{e}$ by generating a grid of models with $M=4.0,8.0,12.0\ M_\mathrm{E}$ and core mass fraction $m_\mathrm{c}/M=0.95,0.90,0.85$, and fitting the pressure and internal energy profiles, finding a best fit with $\gamma_\mathrm{e}=2.957$, and rounding to $\gamma_\mathrm{e}=3$.
We also tried a fit allowing for a temperature dependence on the polytropic term, \begin{equation}p=K_\mathrm{e}\rho^{\gamma_\mathrm{e}}T^{\gamma_\mathrm{T}}+\frac{\rho k_\mathrm{B}T}{\mu m_\mathrm{H}},\end{equation} finding a best fit of $\gamma_\mathrm{T}=-0.0047$, indicating little to no sensitivity to temperature.
Figures~\ref{Fig:FitEoSp} and \ref{Fig:FitEoSu} show the fit for $\gamma_\mathrm{e}=3$ and $\gamma_\mathrm{T}=0,$ which has reasonable agreement for both the pressure and internal energy of all 9 models.
Table~\ref{TBL:Kgas} shows $K_\mathrm{e}$ for each model, and we see that the values fall within a relatively small range.
We also empirically fit $\beta_i$ for each particle to ensure consistency between the internal energy and pressure profiles, where $1.5<\beta_i<2.5$, consistent with an envelope made up of mostly diatomic gas near the surface and monatomic gas at higher pressures.

\begin{deluxetable}{cccc}
\tablewidth{5.0cm}
\tabletypesize{\footnotesize}
\tablecolumns{3}
\tablecaption{$K_\mathrm{e}$ as a function of planet mass and gas-mass fraction \label{TBL:Kgas}}
\tablehead{
    \colhead{$M\ [M_\oplus]$} & \colhead{$m_\mathrm{c}/M$} & \colhead{$K_\mathrm{e}\ [\mathrm{cm}^8\mathrm{g}^{-2}\mathrm{s}^{-2}]$}}
\startdata
$ 4.0  $ & $ 0.85 $ & $ 2.775\times10^{12} $\\
$ 4.0  $ & $ 0.90 $ & $ 2.751\times10^{12} $\\
$ 4.0  $ & $ 0.95 $ & $ 2.766\times10^{12} $\\
$ 8.0  $ & $ 0.85 $ & $ 2.742\times10^{12} $\\
$ 8.0  $ & $ 0.90 $ & $ 2.703\times10^{12} $\\
$ 8.0  $ & $ 0.95 $ & $ 2.721\times10^{12} $\\
$ 12.0 $ & $ 0.85 $ & $ 2.569\times10^{12} $\\
$ 12.0 $ & $ 0.90 $ & $ 2.633\times10^{12} $\\
$ 12.0 $ & $ 0.95 $ & $ 2.659\times10^{12} $\\
\enddata

\tablenotetext{1}{$K_\mathrm{e}$ for our grid of planet-models generated by {\it MESA}.}
\end{deluxetable}

During the calculations we allow the temperature and entropic constants to evolve, calculating pressure as a function of internal energy for pure iron-core particles, \begin{equation}\label{EQ:coreEoS_SPH}p_i=\frac{u_i(\gamma_\mathrm{c}-1)\rho_i\left(1-\frac{\rho_\mathrm{c}^\prime}{\rho_i}\right)^{\gamma_\mathrm{c}}}{{}_2F_1\left(1-\gamma_\mathrm{c},-\gamma_\mathrm{c},2-\gamma_\mathrm{c},\frac{\rho_\mathrm{c}^\prime}{\rho_i}\right)},\end{equation}
pure silicate-mantle particles, \begin{equation}\label{EQ:mantleEoS_SPH}p_i=\frac{u_i(\gamma_\mathrm{m}-1)\rho_i\left(1-\frac{\rho_\mathrm{m}^\prime}{\rho_i}\right)^{\gamma_\mathrm{m}}}{{}_2F_1\left(1-\gamma_\mathrm{m},-\gamma_\mathrm{m},2-\gamma_\mathrm{m},\frac{\rho_\mathrm{m}^\prime}{\rho_i}\right)},\end{equation}
and pure gas particles, \begin{equation}\label{EQ:gasEoS_SPH}p_i=\frac{\rho_iu_i}{\beta_i}+\frac{K_\mathrm{e}\rho_i^{\gamma_\mathrm{e}-1}}{\gamma_\mathrm{e}-1}\left(1-\frac{\rho_i}{\beta_i}\right).\end{equation}

\subsubsection{Resolving Particles at the Interfaces}
Within the core, the density at the interface between the iron core and the silicate mantle is discontinuous, being $\rho_\mathrm{c}$ on the inside of the interface and $\rho_\mathrm{m}$ on the outside.
Due to the density discontinuity, particles near this interface have densities significantly different than the input profile values and we treat them as having a mixed composition, where their equations of state are treated as linear combinations of the iron core and silicate-mantle equations of state, \begin{equation}\label{EQ:uCoreMantleEoS}u_i=x_{\mathrm{c,}i}\frac{c_\mathrm{c}\rho_{\mathrm{c,}i}^{\gamma_\mathrm{c}-1}{}_2F_1\left(1-\gamma_\mathrm{c},-\gamma_\mathrm{c},2-\gamma_\mathrm{c},\frac{\rho^\prime_\mathrm{c}}{\rho_{\mathrm{c},i}}\right)}{\gamma_\mathrm{c}-1}+(1-x_{\mathrm{c,}i})\frac{c_\mathrm{m}\rho_{\mathrm{m},i}^{\gamma_\mathrm{m}-1}{}_2F_1\left(1-\gamma_\mathrm{m},-\gamma_\mathrm{m},2-\gamma_\mathrm{m},\frac{\rho_\mathrm{m}^\prime}{\rho_{\mathrm{m},i}}\right)}{\gamma_\mathrm{m}-1},\end{equation} where $c_\mathrm{c},$ $\rho^\prime_\mathrm{c},$ $\gamma_\mathrm{c},$ $c_\mathrm{m},$ $\rho^\prime_\mathrm{m},$ and $\gamma_\mathrm{m}$ are the fit values for $Fe$ and $MgSi0_3$ respectively (converted from values in Table~3 in \citealt{2007ApJ...669.1279S}), $x_{\mathrm{c,}i}$ is is the mass-fraction of particle $i$ that is iron core, determined during the initial calculation of particle density, $\rho_i$, and smoothing length, $h_i$, by enforcing pressure continuity, \begin{equation}\label{EQ:pressure_continuity}c_\mathrm{c}(\rho_{\mathrm{c},i}-\rho^\prime_\mathrm{c})^{\gamma_\mathrm{c}}=c_\mathrm{m}(\rho_{\mathrm{m},i}-\rho^\prime_\mathrm{m})^{\gamma_\mathrm{m}},\end{equation} and mass conservation, \begin{equation}\rho_i=\begin{cases}
\frac{\rho(r_i)\rho_{\mathrm{m},i}}{x_{\mathrm{c,}i}\rho_{\mathrm{m},i}+(1-x_{\mathrm{c,}i})\rho(r_i)},		& r_i\le R\\
\frac{\rho_{\mathrm{c},i}\rho(r_i)}{x_{\mathrm{c,}i}\rho(r_i)+(1-x_{\mathrm{c,}i})\rho_{\mathrm{c},i}},		& r_i>R, \end{cases}\end{equation} where $\rho_{\mathrm{c},i}$ is the density of the core component of the particle, $\rho_{\mathrm{m},i}$ is the density of the mantle component of the particle, $\rho(r_i)$ is the initially assigned density, $r_i$ is the radial distance of particle $i$, and $R$ is the radial distance of the interface.
We assume that, for particles placed initially within the iron core, $\rho_{\mathrm{c},i}=\rho(r_i)$, and for particles placed initially within the silicate mantle, $\rho_{\mathrm{m},i}=\rho(r_i)$.
After the initial component density values are assigned, we redistribute the internal energy using \eqref{EQ:uCoreMantleEoS}.
During the calculations we solve for $\rho_{\mathrm{c},i}$ and $\rho_{\mathrm{m},i}$ as a function of $u_i$, $x_{\mathrm{c},i}$, and $\rho_i$ by enforcing pressure continuity, \begin{equation}\frac{u_{\mathrm{c},i}(\gamma_\mathrm{c}-1)\rho_{\mathrm{c},i}\left(1-\frac{\rho_\mathrm{c}^\prime}{\rho_{\mathrm{c},i}}\right)^{\gamma_\mathrm{c}}}{{}_2F_1\left(1-\gamma_\mathrm{c},-\gamma_\mathrm{c},2-\gamma_\mathrm{c},\frac{\rho_\mathrm{c}^\prime}{\rho_{\mathrm{c},i}}\right)} = \frac{u_{\mathrm{m},i}(\gamma_\mathrm{m}-1)\rho_{\mathrm{m},i}\left(1-\frac{\rho_\mathrm{m}^\prime}{\rho_{\mathrm{m},i}}\right)^{\gamma_\mathrm{m}}}{{}_2F_1\left(1-\gamma_\mathrm{m},-\gamma_\mathrm{m},2-\gamma_\mathrm{m},\frac{\rho_\mathrm{m}^\prime}{\rho_{\mathrm{m},i}}\right)},\end{equation} and mass conservation, \begin{equation}\label{EQ:MassConservation}\rho_i=\frac{\rho_{\mathrm{c},i}\rho_{\mathrm{m},i}}{x_{\mathrm{c,}i}\rho_{\mathrm{m},i}+(1-x_{\mathrm{c,}i})\rho_{\mathrm{c},i}},\end{equation} where $u_{\mathrm{c},i}$ and $u_{\mathrm{m},i}$ are the internal energies of the core and mantle components, \begin{equation}u_{\mathrm{c},i} = u_i\frac{u_\mathrm{c,0}}{u_\mathrm{0}},\end{equation} \begin{equation}u_{\mathrm{m},i} = u_i\frac{u_\mathrm{m,0}}{u_\mathrm{0}},\end{equation} and $u_\mathrm{0}$, $u_\mathrm{c,0}$, and $u_\mathrm{m,0}$ are the initial values of the internal energies found from \eqref{EQ:uCoreMantleEoS}.

At the interface between the mantle's surface and the gas envelope, the density drops from $\rho_\mathrm{s}$ to $\rho_\mathrm{e}$, as determined by the input profile.
Due to the density discontinuity, particles near this interface have densities significantly different the input profile values and we treat them as having a mixed composition, where their equations of state are treated as linear combinations of the gas and silicate-mantle equations of state, \begin{equation}\label{EQ:uMantleGasEoS}u_i=x_{\mathrm{m},i}\frac{c_\mathrm{m}\rho_{\mathrm{m},i}^{\gamma_\mathrm{m}-1}{}_2F_1\left(1-\gamma_\mathrm{m},-\gamma_\mathrm{m},2-\gamma_\mathrm{m},\frac{\rho_\mathrm{m}^\prime}{\rho_{\mathrm{m},i}}\right)}{\gamma_\mathrm{m}-1}+(1-x_{\mathrm{m},i})\left(\frac{K_\mathrm{e}\rho_\mathrm{e}^{\gamma_\mathrm{e}-1}}{\gamma_\mathrm{e}-1}+\frac{\beta_ik_\mathrm{B}T_i}{\mu_im_H}\right),\end{equation} where $x_{\mathrm{m,}i}$ is is the mass-fraction of particle $i$ that is silicate mantle, determined during the initial calculation of particle density, $\rho_i$, and smoothing length, $h_i$, by enforcing pressure continuity, \begin{equation}\label{EQ:pressure_continuity2}c_\mathrm{m}(\rho_{\mathrm{m},i}-\rho^\prime_\mathrm{m})^{\gamma_\mathrm{m}} = K_\mathrm{e}\rho_i^{\gamma_\mathrm{e}}+\frac{\rho_ik_\mathrm{B}T_i}{\mu_im_H},\end{equation} and mass conservation, \begin{equation}\rho_i=\begin{cases}
\frac{\rho(r_i)\rho_{\mathrm{e},i}}{x_{\mathrm{m},i}\rho_{\mathrm{e},i}+(1-x_{\mathrm{m},i})\rho(r_i)},		& r_i\le R\\
\frac{\rho_{\mathrm{m},i}\rho(r_i)}{x_{\mathrm{m},i}\rho(r_i)+(1-x_{\mathrm{m},i})\rho_{\mathrm{m},i}},		& r_i>R,\end{cases}\end{equation} where $\rho_{\mathrm{m},i}$ is the density of the mantle component of the particle and $\rho_{\mathrm{e},i}$ is the density of the gas component of the particle.
We assume that, for particles placed initially within the silicate mantle, $\rho_{\mathrm{m},i}=\rho(r_i)$, and for particles placed initially within the gas envelope, $\rho_{\mathrm{e},i}=\rho(r_i)$.
After the initial density values are assigned we redistribute the internal energy using \eqref{EQ:uMantleGasEoS}.
During the calculations we solve for $\rho_{\mathrm{m},i}$ and $\rho_{\mathrm{e},i}$ as a function of $u_i$, $x_{\mathrm{m},i}$, and $\rho_i$ by enforcing pressure continuity,\begin{equation}\frac{u_{\mathrm{m},i}(\gamma_\mathrm{m}-1)\rho_{\mathrm{m},i}\left(1-\frac{\rho_\mathrm{m}^\prime}{\rho_{\mathrm{m},i}}\right)^{\gamma_\mathrm{m}}}{{}_2F_1\left(1-\gamma_\mathrm{m},-\gamma_\mathrm{m},2-\gamma_\mathrm{m},\frac{\rho_\mathrm{m}^\prime}{\rho_{\mathrm{m},i}}\right)} = \frac{u_{\mathrm{e},i}\rho_{\mathrm{e},i}}{\beta_i}+K_\mathrm{e}\rho_{\mathrm{e},i}^{\gamma_\mathrm{e}}\left(1-\frac{1}{(\gamma_\mathrm{e}-1)\beta_i}\right)\end{equation} and mass conservation, \begin{equation}\label{EQ:MassConservation2}\rho_i=\frac{\rho_{\mathrm{m},i}\rho_{\mathrm{e},i}}{x_{\mathrm{m,}i}\rho_{\mathrm{e},i}+(1-x_{\mathrm{m,}i})\rho_{\mathrm{m},i}},\end{equation} where $u_{\mathrm{m},i}$ and $u_{\mathrm{e},i}$ are the initial internal energies of the mantle and gas components, \begin{equation}u_{\mathrm{m},i} = u_i\frac{u_\mathrm{m,0}}{u_\mathrm{0}},\end{equation} \begin{equation}u_{\mathrm{e},i} = u_i\frac{u_\mathrm{e,0}}{u_\mathrm{0}},\end{equation} and $u_\mathrm{0}$, $u_\mathrm{m,0}$, and $u_\mathrm{e,0}$ are the initial values of the internal energies found from \eqref{EQ:uMantleGasEoS}.

\begin{figure}[htp]
\begin{center}
\begin{tabular}{cc}
\includegraphics[width=8.0cm]{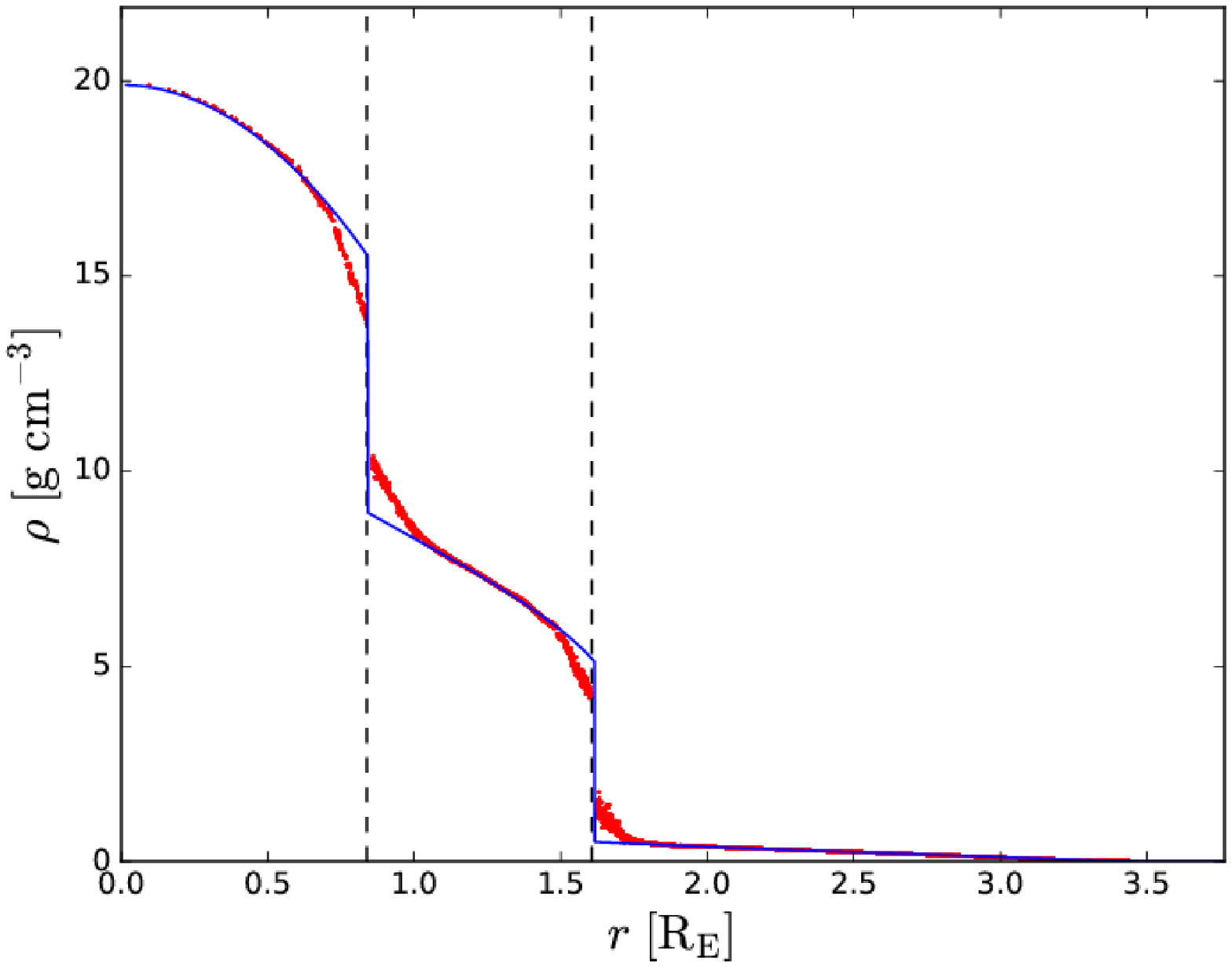}
\includegraphics[width=8.0cm]{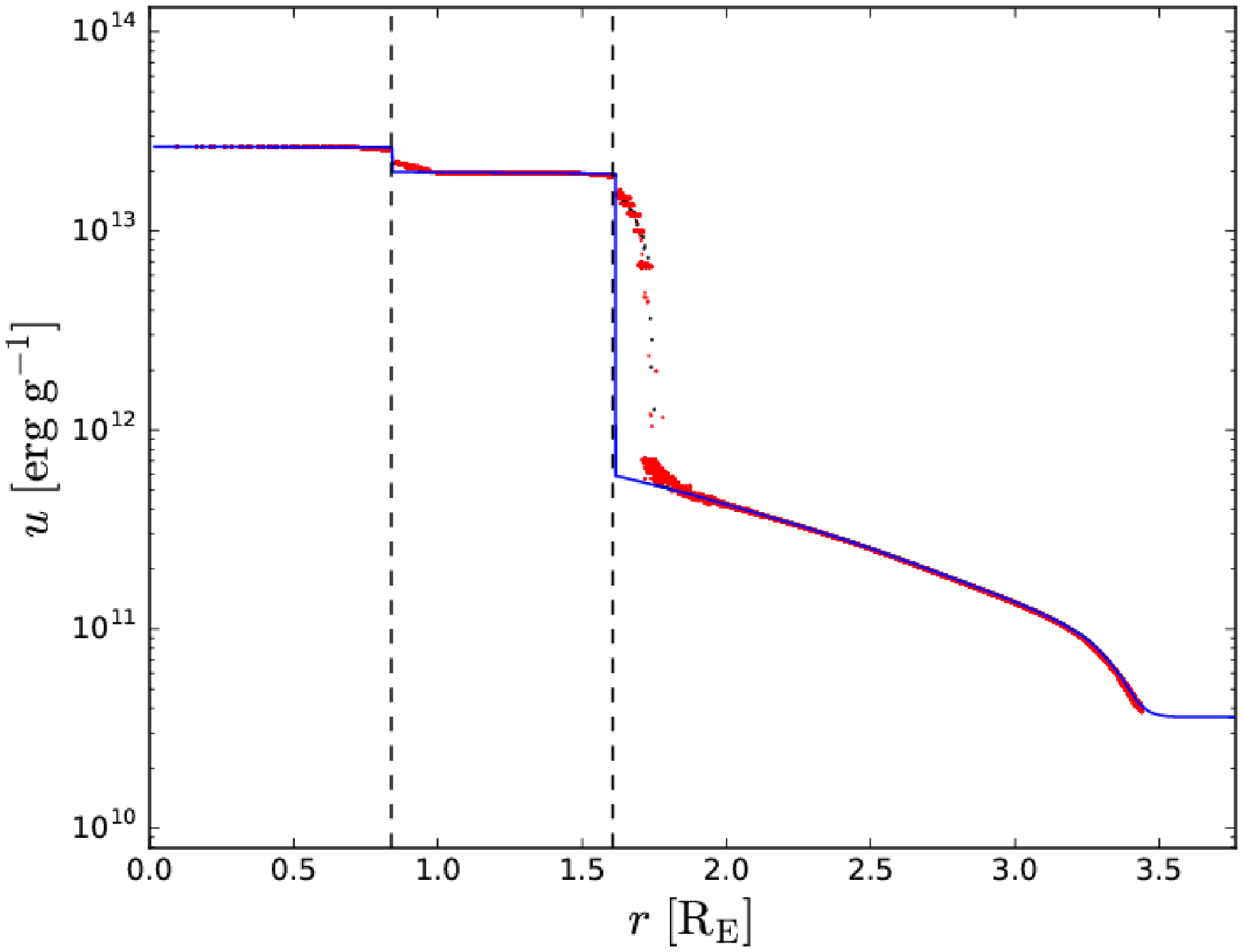}\\
\includegraphics[width=8.0cm]{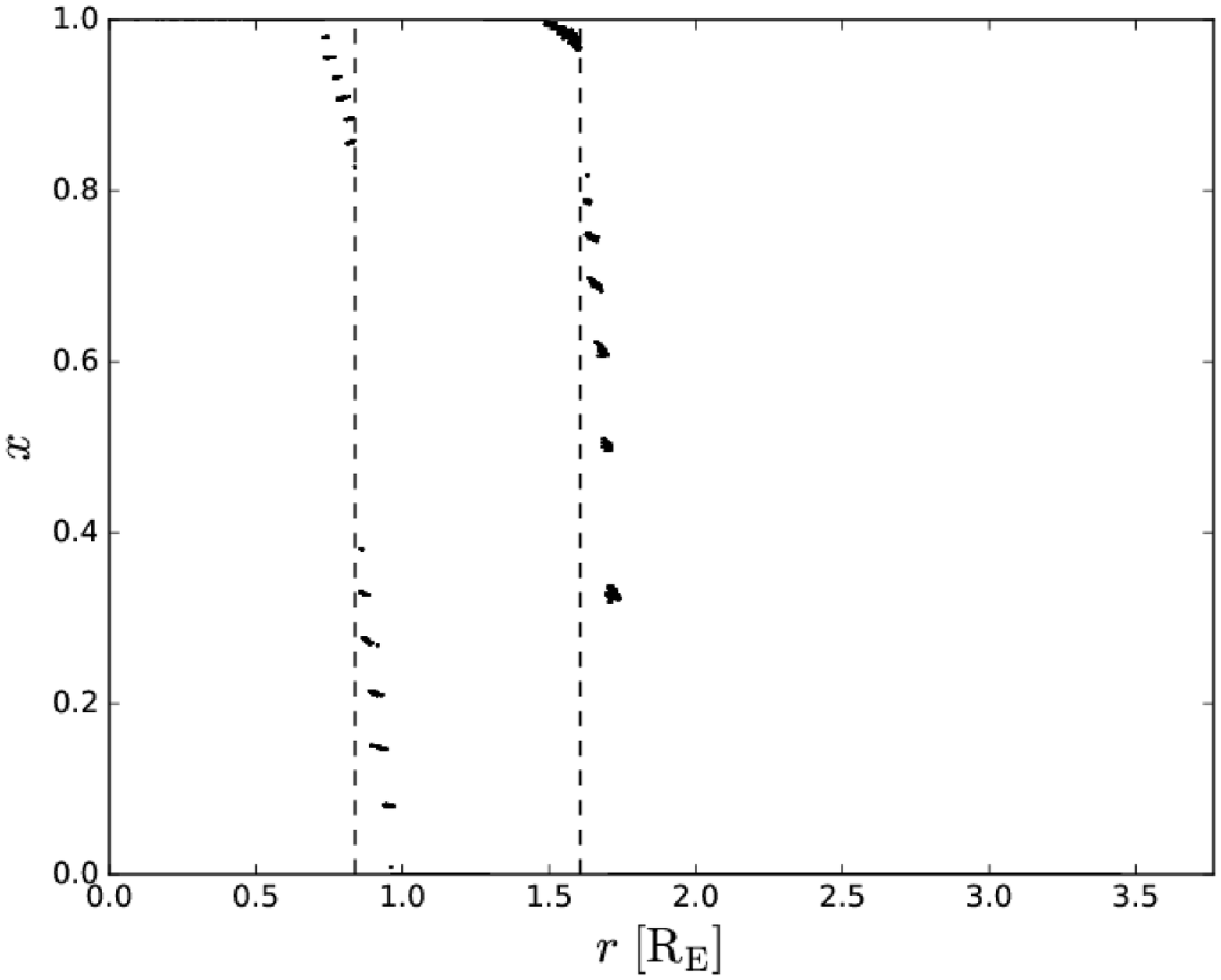}
\includegraphics[width=8.0cm]{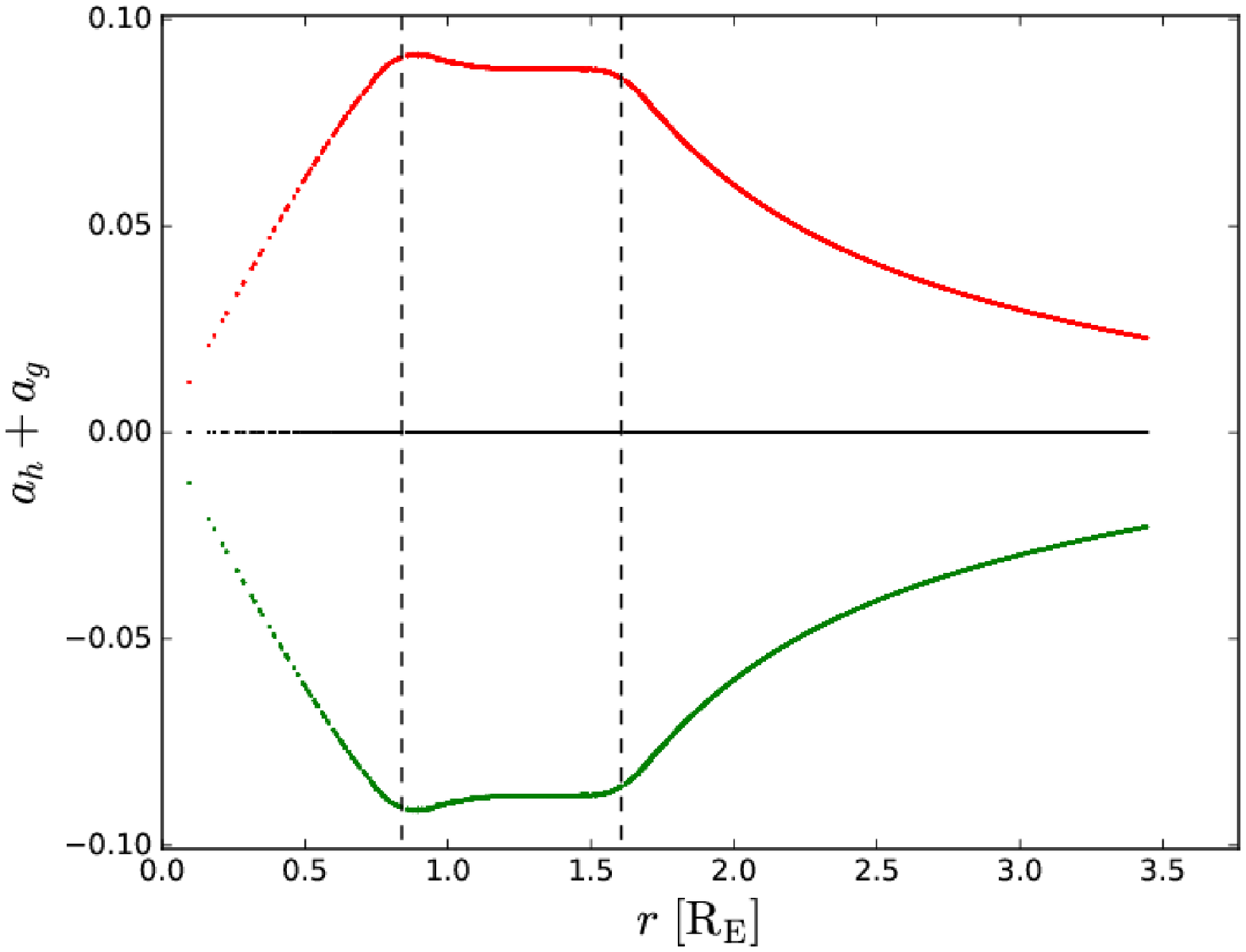}\\
\end{tabular}
\end{center}
\caption{Radial profiles at the end of 2000 dynamical times for an isolated model of a Kepler-11e analog, where we enable a relaxation force (see \citealt{2015ApJ...806..135H} for details) for the first 100 dynamical times.
In each plot we mark the transitions from core to mantle and from mantle to gas envelope (vertical dashed lines).
Density (top left) and internal energy (top right) profiles for the input profile (blue), generated using {\it MESA} and the prescriptions from \citet{2007ApJ...669.1279S}, the initial values assigned to each particle (black), and the values after 2000 dynamical times (red).
Particle composition, $x$, profile (bottom left) after 2000 dynamical times.
Equilibrium profile (bottom right) after 2000 dynamical times, showing the radial acceleration from the hydrostatic force (red), the gravitational force (green), and the total force (black).
These figures show how we assigned the particle composition and redistributed internal energy based on the smoothed densities, where $x$ represents the mass-fraction of the particle belonging to the heavier composition near the interface.
The model relaxes into a very stable hydrostatic equilibrium, stable for at least the timescales used for the dynamical calculations.}
\label{Fig:Radial_Profiles}
\end{figure}

Figure~\ref{Fig:Radial_Profiles} compares the planet profiles generated using {\it MESA} and the relaxed SPH planet models created with {\it StarSmasher} after 2000 dynamical times, and we see that the initial profile is maintained reasonably well throughout the relaxation run.
The planet models are very stable at the end of the relaxation; the radial acceleration on the particles is near zero throughout the profile.

\subsection{StarSmasher}
We use an SPH code, {\it StarSmasher} (previously {\it StarCrash}, originally developed by \citealt{1991PhDT........11R} and later updated as described in \citealt{1999JCoPh.152..687L} and \citealt{2000PhRvD..62f4012F}) for the hydrodynamic calculations.
The code now implements variational equations of motion and libraries to calculate the gravitational forces between particles using direct summation on NVIDIA graphics cards as described in \citet{2010MNRAS.402..105G}.
Using a direct summation instead of a tree-based algorithm for gravity increases the accuracy of the gravity calculations at the cost of speed \citep{2010ProCS...1.1119G}.
The code uses a cubic spline \citep{1985A&A...149..135M} for the smoothing kernel and an artificial viscosity prescription coupled with a Balsara Switch \citep{1995JCoPh.121..357B} to prevent unphysical inter-particle penetration, specifically described in \citet{2015ApJ...806..135H}.
StarSmasher is available at https://jalombar.github.io/starsmasher/.

\end{document}